\documentclass[aps,10pt,prc,tightenlines,twocolumn,twoside,showpacs,nofootinbib,superscriptaddress]{revtex4-1}
\usepackage{graphicx,amsmath,amssymb}
\usepackage{hyperref}
\usepackage{breakurl}
\usepackage{color}
\usepackage{verbatim}
\usepackage{multirow}
\usepackage{tabularx}
\usepackage{slashed}
\usepackage{soul}

\newcommand{\eg}{\textit{e.g.}}  

\newcommand{\ie}{\textit{i.e.}}

\newcommand{\be}{\begin {equation}}
\newcommand{\ee}{\end{equation}}
\newcommand{\bi}{\begin{itemize}}
\newcommand{\ei}{\end{itemize}}
\newcommand{\bea}{\begin {eqnarray}}
\newcommand{\eea}{\end{eqnarray}}
\newcommand{\braket}[2]{\bra{#1}\,#2\rangle} 
\newcommand{\bra}[1]{\langle\,#1\,|}          
\newcommand{\ket}[1]{|\,#1\,\rangle}          
\newcommand{\ud}{\mathrm{d}}

\newcommand{\LCm}{{\scriptscriptstyle -}} 
\newcommand{\LCp}{{\scriptscriptstyle +}}
\newcommand{\LCpm}{{\scriptscriptstyle \pm}}
\newcommand{\LCmp}{{\scriptscriptstyle \mp}}

\newcommand{\LCperp}{{\scriptscriptstyle \perp}}

\begin{document}

\title{Scattering in Time-dependent Basis Light-Front Quantization}

\author{Xingbo Zhao}
\email{xbzhao@iastate.edu}
\affiliation{Department of Physics and Astronomy, Iowa State University, Ames, Iowa 50011, USA}

\author{Anton Ilderton} 
\email{anton.ilderton@chalmers.se}
\affiliation{Department of Applied Physics, Chalmers, SE-41296 G\"oteborg, Sweden}
\author{Pieter Maris}
\email{pmaris@iastate.edu}
\affiliation{Department of Physics and Astronomy, Iowa State University, Ames, Iowa 50011, USA}

\author{James P. Vary} 
\email{jvary@iastate.edu}
\affiliation{Department of Physics and Astronomy, Iowa State University, Ames, Iowa 50011, USA}

\date{\today}

\begin{abstract}
We introduce a nonperturbative, first principles numerical approach for solving time-dependent problems in quantum field theory, using light-front quantization. As a first application we consider QED in a strong background field, and the process of non-linear Compton scattering in which an electron is excited by the background and emits a photon. We track the evolution of the quantum state as a function of time. Observables, such as the invariant mass of the electron-photon pair, are first checked against results from perturbation theory, for suitable parameters. We then proceed to a test case in the strong background field regime and discuss the various nonperturbative effects revealed by the approach.
\end{abstract}


\maketitle

\section{Introduction}
\label{sec_intro}

Treating quantum field theory in the nonperturbative regime remains a significant challenge.  ``Basis Light-Front Quantization'' (BLFQ)~\cite{Vary:2009gt},
which adopts light-front quantization and the Hamiltonian formalism, offers a first-principles approach to nonperturbative quantum field theory (QFT)~\cite{Vary:2009gt,Honkanen:2010rc}. Diagonalization of the full Hamiltonian of the quantum field theory yields the physical eigenvalues and eigenvectors of the mass eigenstates. This approach offers new insights into bound state properties and scattering processes~\cite{Brodsky:1997de} as well as opportunities to address many outstanding puzzles in nuclear and particle physics~\cite{Brodsky:2011vc,Brodsky:2012rw}. 

The BLFQ approach is real-time (as opposed to imaginary-time, as normally used in lattice-QFT, see though~\cite{Hebenstreit:2013qxa}) and therefore naturally applicable to time-dependent problems. There is currently much interest in gauge theories with an explicit time dependence introduced by a background field, in particular QED in ultra-intense laser fields \cite{Heinzl:2008an,DiPiazza:2011tq} and QCD in strong magnetic fields \cite{Chernodub:2011mc,Bali:2011qj,Basar:2011by,Tuchin:2012mf}. In both cases, the greatest interest lies in the case for which the fields are strong enough to require a nonperturbative treatment and this motivates the approach we present here.

In this paper we introduce {\it time-dependent} Basis Light-Front Quantization (tBLFQ), which is an extension of BLFQ to time-dependent problems in quantum field theory. In this approach, BLFQ provides the eigenstates of the time-independent part of the Hamiltonian. We then solve for the time evolution of a chosen initial state under the influence of an applied background field, which is introduced through explicitly time-dependent interaction terms in the Hamiltonian. Although we treat a specific application in the present work, the method is more generally applicable to time evolution even in the absence of external fields where one is simply following the evolution of a chosen non-stationary state of the system.

In this paper  we will apply tBLFQ to ``strong field QED'', in which the background field models the high-intensity fields of modern laser systems. Such light sources now routinely reach intensities of $10^{22}$ W/cm$^2$, and there is ongoing research into using intense lasers to investigate previously unmeasured effects such as vacuum birefringence~\cite{Heinzl:2006xc,King} and Schwinger pair production~\cite{Dunne:2008kc}. Within this research field, the use of large-scale numerical codes, based on kinetic models, is becoming increasingly popular~\cite{Fedotov:2010ja, Sokolov:2010am,Elkina:2010up, King:2013zw}. The two main advantages of such approaches are that they are real-time, and that huge numbers of particles can be treated via particle-in-cell (PIC) simulations. However, there exists no first-principles derivation of the required kinetic equations from QED. Consequently, this approach is based on a forced welding of classical and quantum theories, in which particles and photons are treated as classical ballistic objects, and QED cross sections are added by hand to model instantaneous collisions. This leads to problems with double-counting and the inclusion of higher-order processes.

Here, we consider an alternative approach. We restrict ourselves to low numbers of particles, but we perform a fully quantum and real-time calculation within QED. Specifically, we will study ``non-linear Compton scattering'' (nCs), in which an electron is excited by a background field and emits a photon \cite{Nikishov:1963,Nikishov:1964a}. This is one of the simplest background field processes, as there are no thresholds to overcome as in, say, pair creation. We note that light-front quantization is the natural setting for this investigation \cite{Neville:1971zk, Ilderton:2012qe}, since lasers have inherently ``light-front'' properties:  all photons propagate on the light-front.

This paper is organized as follows. We provide the background to our approach in Sec.~\ref{sec:bg}, followed by details of the BLFQ method in Sec.~\ref{sec:BLFQ}.  We then introduce tBLFQ in Sec.~\ref{sec:tBLFQ} and provide illustrative numerical results for our first application to non-linear Compton scattering in Sec.~\ref{sec:result}.  We present our conclusions and outlook in Sec.~\ref{sec:concl}. The Appendices contain a number of useful details.

\section{Background}
\label{sec:bg}
Our approach is based on light-front quantization, and on a previously developed method called BLFQ~\cite{Vary:2009gt,Honkanen:2010rc}.  We begin here with a brief review of relevant aspects of the light-front formalism~\cite{Brodsky:1997de, Heinzl:2000ht}, and an outline, in terms of textbook methods, of the calculation which we wish to perform.

Physical processes in light-front dynamics are described in terms of light-front coordinates ($x^+$, $x^1$, $x^2$, $x^-$), in which $x^+$=$x^0$+$x^3$ plays the role of time. Hence, quantization surfaces are null hyperplanes given by $x^+$=constant, and on which initial conditions are specified. $x^-$=$x^0{-}x^3$ is the ``longitudinal'' direction, and the remaining two spatial directions are called ``transverse'', $x^\perp$=\{$x^1$, $x^2$\}. The evolution of quantum states is governed as usual by the Schr\"odigner equation, which in light-front quantization takes the form
\begin{align}
\label{wave-eq}
	i\frac{\partial}{\partial x^+}\ket{\psi;x^+}= \frac{1}{2}P^-(x^+)\ket{\psi;x^+} \;,
\end{align}
where $\ket{\psi;x^+}$ is the (Schr\"odinger picture)  state at light-front time $x^+$ and $P^-$ is the light-front Hamiltonian. Our Hamiltonian will contain two parts; $P^-_\text{QED}$ which is the full light-front Hamiltonian of QED, and $V$ which contains interaction terms introduced by a background field, so
\begin{align}
	\label{H_interact}
	P^-(x^+)=P^-_\text{QED}+V(x^+) \;.
\end{align}
$V$ contains, in general, an explicit time dependence. It is therefore natural to use an interaction picture, but we must immediately stress two things: first, we are {\it not} using the usual ``free + interacting'' split of the Hamiltonian and, second, we are not working in perturbation theory.  Instead, the full QED Hamiltonian $P_\text{QED}^-$ replaces the customary ``free'' Hamiltonian, and $V$ is naturally the interaction term. The interaction picture states are then defined by
\be
	\ket{\psi;x^+}_I = e^{\tfrac{i}{2}P^-_\text{QED}x^+}\ket{\psi;x^+} \;,
\ee
(since $P^-_\text{QED}$ is time-independent), and obey
\be\label{Schro-int}
	i\frac{\partial}{\partial x^+}\ket{\psi;x^+}_I= \frac{1}{2}V_I(x^+)\ket{\psi;x^+}_I \;,
\ee
in which $V_I$, ``the interaction Hamiltonian in the interaction picture'', is
\be
	V_I(x^+) = e^{\tfrac{i}{2}P^-_\text{QED}x^+}V(x^+)e^{-\tfrac{i}{2}P^-_\text{QED}x^+} \;.
\ee
The formal solution to (\ref{Schro-int}) is
\begin{align}
	\label{i_evolve}
	\ket{\psi;x^+}_I  &= \mathcal{T}_+ \exp\bigg(-\frac{i}{2}\int\limits_0^{x^+} V_I\bigg)\ket{\psi;0}_I \;,
\end{align}
where $\mathcal{T}_\LCp$ is light-front time ordering. Now let us imagine that we could ``solve'' QED and identify the eigenstates and eigenvalues of the theory. Call these $\ket{\beta}$ and $P^-_\beta$ respectively, so
\begin{align}\label{Energi}
	P^-_\text{QED}\ket{\beta} = P^-_\beta \ket{\beta} \;. 
\end{align}
Having these covariant solutions, we would then be interested in the transitions between such states introduced by the background field interactions contained in $V$. We choose the external field (modeling an intense laser) to vanish , $V = 0$, prior to $x^+ = 0$. At $x^+=0$, we expand a chosen initial state as a sum over QED eigenstates:
\begin{align}
	\label{initial_c}
	\ket{\psi;0}_I = \sum\limits_\beta \ket{\beta} c_\beta(0)\;,
\end{align}
where $c_\beta(0)$ is the initial data such that
\be\label{IC}
	c_\beta(0) \equiv \braket{\beta}{\psi;0}_I \;.
\ee
We then expand a solution of the interaction picture state at later times,
\be\label{int-expand}
	\ket{\psi;x^+}_I  := \sum_{\beta} c_\beta(x^+) \ket{\beta},
\ee
in which the coefficients $c_\beta$ characterize the nontrivial part of the state's time evolution induced by the external field. Plugging (\ref{int-expand}) into (\ref{Schro-int}) yields an equation for the $c_\beta$: 
\begin{align}
\label{wave-eq_i}
	 i\frac{\partial c_\beta(x^+)}{\partial x^+} 
	&= \sum_{\beta'} \bra{\beta} \tfrac{1}{2}V_I(x^+) \ket{\beta'} c_{\beta'}(x^+) \nonumber\\
	&\equiv \mathcal{M}_{\beta\beta'}(x^\LCp) c_{\beta'}(x^+).
\end{align}
(Summation notation in the second line.) This is an intractable infinite-dimensional system of coupled differential equations, and it is at this point that one would normally switch to perturbation theory in the interaction~$V$. However, the background fields we wish to treat are strong and therefore not amenable to perturbation theory. We therefore write down the formal solution to (\ref{wave-eq_i}), which is, regarding $c_\beta$ as a column vector and $\mathcal{M}_{\beta\beta'}$ as a matrix, both with infinite dimensions,
\begin{align}
	\label{c_evolve}
	c(x^+) &= \mathcal{T}_+ \exp\bigg(-i \int\limits_0^{x^+} \mathcal{M}\bigg)c(0) \;.
\end{align}
In our approach, BLFQ provides finite dimensional approximate solutions for the eigenstates $\ket{\beta}$. In tBLFQ, the time evolution in (\ref{c_evolve}) is performed numerically, beginning with the initial vector $c(0)$, to find the vector $c(x^\LCp)$. The coefficients $c_\beta(x^\LCp)$ can then be read off, allowing one to reconstruct the evolved state itself from the overlap
\begin{align}
	\label{state_interact_wf}
	c_\beta(x^+)=\braket{\beta}{\psi;x^+}_I \;.
\end{align}
In this way we solve equation (\ref{wave-eq_i}) with initial conditions~(\ref{IC}).

Let us compare the above to the usual calculation of scattering amplitudes in QED. Such amplitudes are based on the split of the QED Hamiltonian into a free particle Hamiltonian, $P^\LCm_\text{free}$, and an interaction.  For the application here, this split produces an interaction that would be the sum of the QED interaction terms, call them $V_\text{Q}$, and the additional interaction terms introduced by the background, $V$.

A scattering calculation would begin with an initial state which is a free particle state $\ket{i}$, prepared at $x^\LCp{=}-\infty$. This state would be evolved through all time using the $S$-matrix operator~\cite{Weinberg},
\begin{align}
	\label{s-matrix}
	S=\lim_{T\to\infty} \mathcal{T}_+ e^{-\frac{i}{2}\int\limits_{-T}^{T} {V_{Q}}_I + V_I}\;.
\end{align}
and projected onto a final state $\ket{f}$, describing free particles at $x^\LCp{=}+\infty$. Thus, one obtains the $S$-matrix element
\be
	S_{fi} = \bra{f} S \ket{i} \;.
\ee
We are also calculating ``scattering amplitudes'', but there are two important differences between our approach and that based on the $S$-matrix. First, we calculate transitions based upon the eigenstate basis of QED (for example physical electrons) rather than between free particle states. Second, and related, we calculate finite-time, rather than asymptotic, transitions between such states. For all times before and after the external field acts on our chosen state, we have, in principle, the full quantum amplitude expressed as a superposition of physical states (mass eigenstates of QED). A specific experimental setup will then project this full amplitude onto states to which that setup is sensitive. 
%

\subsection{Application: Nonlinear Compton Scattering}
\label{ssec_laser}
%
In this paper we apply tBLFQ to the process of single photon emission from an electron accelerated by a background field. Taking the background to model an intense laser, this process often goes by the name ``non-linear Compton scattering'' and is well-studied in plane wave backgrounds~\cite{Boca:2009zz,Heinzl:2009nd,Seipt:2010ya, Mackenroth:2010jr}. An appropriate experimental setup would see the (almost head on) collision of an electron with the laser, and the subsequent measurement of either the emitted photon~\cite{Harvey:2012ie} or electron~\cite{Boca:2012pz} spectra.

\begin{figure}[!t]
\centering
\centering\includegraphics[width=0.45\textwidth]{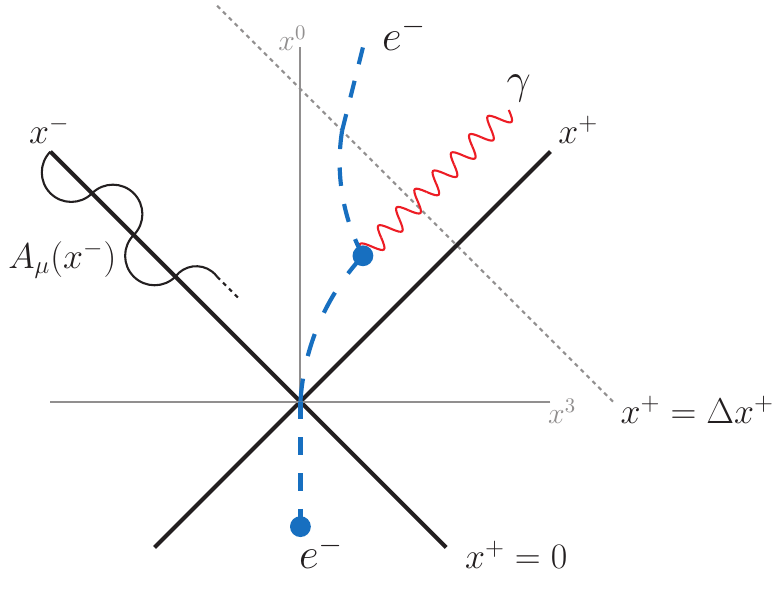}
\caption{An illustration of non-linear Compton scattering. An electron enters a laser field, is accelerated, and emits a photon. After emission the electron can be further accelerated until it leaves the field.}
\label{fig:nCs}
\end{figure}

We begin with an electron at light-front time $x^+$=0 when it first encounters the laser field. The electron may be both accelerated (invariant mass unchanged but 4-vector altered) and excited (invariant mass changed) by the laser field. Excitation produces electron--photon final states. After time $\Delta x^+$ the background field switches off and no further acceleration or excitation may occur. This setup is sketched in Fig.~\ref{fig:nCs} for two of the four dimensions in the problem. The natural question to ask is how the quantum states of the electron and (emitted) photon fields evolve with light-front time $x^+$, and this will indeed be studied below.

While, in principle, there is nothing to stop us including arbitrarily complex background fields, as a first step we consider a simple model. The background is turned on only for finite light-front time $\Delta x^\LCp$, during which it is independent of $x^\LCp$ but inhomogeneous in $x^-$,
\begin{align}
\label{laser-profile}
	 e\mathcal{A}^-(x^-)	&=2 m_e a_0 \cos{(l_\LCm x^-)}\\\nonumber &=m_e a_0 \left[\exp{(il_\LCm x^-)}+\exp{(-il_\LCm x^-)}\right].
\end{align}
where $e$ is the electron charge and $m_e$ is the electron mass. We have written out the exponential form of cosine to highlight that the field both ``pushes'' and ``pulls'' particles in the longitudinal direction. This field has periodic structure in the longitudinal direction with frequency $\omega = l_\LCm$ and the dimensionless parameter $a_0$ measures the field strength in relativistic units, $a_0 = eE /m_e\omega$. ($a_0=1$ corresponds to an intensity of $\sim 10^{18}$ W/cm$^2$ at optical frequency~\cite{Heinzl:2008an}.) It is uniform in the transverse plane, as for plane waves, but unlike plane waves is longitudinally polarized. The profile (\ref{laser-profile}) describes, in the lab frame, a beam of finite duration $\sqrt{2}\Delta x^+$ propagating along the $x^3$ direction. Classically, such a field accelerates charges in the $x^-$ ($x^3$) direction as time $x^+$ ($x^0$) evolves. The accelerated charges subsequently radiate, see Fig.~\ref{fig:nCs}, and it is the quantum version of this radiation which we will investigate below.

Note that (\ref{laser-profile}) does not obey Maxwell's equations in vacuum. This is not an issue for us since we are interested here not in phenomenology but in a first demonstration of the framework of tBLFQ. Whether the background obeys Maxwell or not has no impact on our methods. With future developments of our formalism in mind, we note that a simple background field model obeying Maxwell would be a plane wave. However, it is also common to consider time-dependent electric fields, which do not obey Maxwell, as models of the focus of counter-propagating pulses~\cite{Dunne:2008kc}. Insisting on background field profiles which are both realistic (finite energy, pulsed in all four dimensions) and obey Maxwell's equations is a challenge, as very few such solutions exist in closed form. An exception is given in~\cite{IVAN}, and while there is nothing to stop us including such backgrounds in principle, doing so goes somewhat beyond the initial ``proof-of-concept" presented here.

\section{Basis Light-front Quantization (BLFQ)}
\label{sec:BLFQ}

We are interested in how eigenstates of the {\it full} QED Hamiltonian $P^-_\text{QED}$ evolve due to interactions with a background field. (This is analogous to, but clearly not the same as, studying transitions between bare states induced by perturbative QED interactions.) To begin, we must therefore find the eigenstates of QED, for which we must adopt an approximation.

The method we use to construct the approximate eigenstates is Basis Light-front Quantization, or BLFQ~\cite{Vary:2009gt,Honkanen:2010rc}. This is a numerical method for calculating the spectrum of a Hamiltonian, using light-front quantization. The idea of finding, for example, the bound state spectrum via diagonalization of the Hamiltonian has a long history~\cite{Brodsky:1997de}. One well-known approach is discretized light-cone quantization~\cite{Brodsky:1997de,Pauli:1985pv,Pauli:1985ps,Tang:1991rc}, on which BLFQ is in part based. The idea behind BLFQ, and its main advantage, is that its adopted basis should have the same symmetries as the full QED or QCD Hamiltonian. (BLFQ was initially designed for QCD~\cite{Vary:2009gt} and is supported by successful anti-de Sitter QCD methods~\cite{deTeramond:2008ht}.) This basis is therefore not the usual basis of momentum states. Usually, the more symmetries the basis captures, the less computational effort is needed for the solutions to reflect those symmetries. Because of this, BLFQ achieves an accurate representation of the Hamiltonian using available computational resources. The construction of the BLFQ basis therefore begins with symmetries of the light-front Hamiltonian.

\subsection{Basis construction} \label{3a}
The derivation of the light-front QED Hamiltonian (in the presence of background fields) and a list of relevant mutually commuting operators, may be found in Appendix~\ref{hami_full}. We do not need the detailed form of these operators in order to discuss the three symmetries directly encoded in the BLFQ basis.  A fourth symmetry, transverse boost invariance,
(also referred to as transverse Galilei invariance, \cite{Brodsky:1997de,Heinzl:2008an}) is discussed separately below as it is not encoded directly in the BLFQ basis but is easily accessible with the employed transverse basis.

The three directly encoded symmetries are 1) Translational symmetry in the longitudinal $x^-$ direction. The longitudinal momentum operator, $P^+$, therefore commutes with the Hamiltonian $[P^-_\text{QED}, P^+]=0$, and total longitudinal momentum is conserved. 2) Rotational symmetry in the transverse plane. This means that the longitudinal projection of angular momentum is conserved, and the corresponding operator $J^3$ obeys $[P^-_\text{QED}, J^3]=0$. The operator $J^3$ can be decomposed into two parts for each particle species,
\be\label{two-part-decomp}
	J^3=J^3_{o}+J^3_{i} \;,
\ee
in which the subscript $``o"$ refers to the longitudinal projection of orbital angular momentum, while subscript $``i"$ refers to the longitudinal projection of the spin angular momentum. This defines the helicity of a particle in light-front dynamics.  3) Charge conservation $[P^-_\text{QED}, Q]=0$, where $Q$ is the charge operator with eigenvalue equal to the net fermion number $N_f$.

The existence of these conserved quantities means that the QED eigenspace can be divided up into ``segments", which are groups of eigenstates with definite eigenvalues\footnote{Throughout, $K$ is a an integer except when an odd number of fermions are present in the basis state for which it is a half-integer. The constant of proportionality is explained below, see equation (\ref{longitudinal_momentum})} $P^+ \propto K$, $J^3=M_j$ and $Q=N_f$.  The full spectrum of QED is the sum of all such segments.

The BLFQ basis is chosen to respect these symmetries. The essential point is that each basis state, call them $\ket{\alpha}$, is an eigenstate of the three operators introduced above, with the eigenvalues,
\be
	\{J^3, P^+, Q\} \ket{\alpha} = \{M_j, K, N_f\}\ket{\alpha} \;.
\ee
Therefore, each state belongs to one and only one segment. As a consequence, the BLFQ basis divides into segments, and the QED Hamiltonian $P^-_\text{QED}$ is accordingly {\it block-diagonal} in the BLFQ basis. As will be outlined below, this structure allows for a large reduction in (numerical) complexity in bound-state calculations.

The BLFQ basis states are built for each Fock-sector (of free-particle states) by allowing the particles to occupy orthonormalized modes of a single-particle basis that facilitates implementation of the full symmetries. The many-particle basis states in each Fock-sector are therefore direct products of single particle states, written $\ket{\bar\alpha}$, so $\ket{\alpha}=\otimes\ket{\bar\alpha}$ in general. It clearly remains to specify the details of the single-particle states.

The single particle basis states are chosen to be two-dimensional harmonic oscillator (``2D-HO'') states in the transverse direction and discretized plane waves in the longitudinal direction.  This is one choice (among many) that facilitates implementation of the symmetries mentioned above.  We note in passing the contrast with treatments of the transverse degrees of freedom in a discretized two-dimensional plane wave basis where the orbital projection symmetry is lost.  We also note the freedom to choose another orthonormal basis in the transverse space using cylindrical coordinates that may be better for some applications.  

Each single particle state carries four quantum numbers,
\begin{align}
    \label{single_particle_labeling}
   \bar\alpha =\{k,n,m,\lambda\} \; .
\end{align}
The first quantum number, $k$, labels the particle's longitudinal momentum. For this degree of freedom we employ the usual plane-wave basis states, \ie\ eigenstates of the free-field longitudinal momentum operator $P^+$, see (\ref{longitudinal_momentum_operator}), with corresponding eigenvalues $p^+$. In this paper, we compactify $x^\LCm$ to a circle of length $2L$. We impose (anti) periodic boundary conditions on (fermions) bosons. As a result, the longitudinal momentum $p^+$ in our basis states takes the discrete values
\begin{align}
    \label{longitudinal_momentum}
    p^+=\frac{2\pi}{L}k
\end{align}
where the dimensionless quantity $k$=1, 2, 3,... for bosons (neglecting the zero mode) and $k=\frac{1}{2}, \frac{3}{2}, \frac{5}{2}$ for fermions.   In particular, we have for the laser $l^\LCp =\tfrac{2\pi}{L}k_\text{las}$ where $k_\text{las}$ is a natural number.  For convenience, throughout this paper we take $L=2\pi\,$MeV$^{-1}$ so that $k$ can be interpreted as the longitudinal momentum in units of MeV.

The next two quantum numbers, $n$ and $m$, label the degrees of freedom in the transverse directions. As mentioned above we take the transverse components of our single particle states to be eigenstates of a 2D-HO which is defined by two parameters, mass $M$ and frequency $\Omega$. (See below for the characteristic scale of the oscillator, which depends only on a combination of these parameters.)  These eigenstates are labelled by the quanta of the radial excitation,  $n$, and the angular momentum quanta, $m$. The eigenstate carrying these numbers has HO eigenenergy
\be\label{ENM}
	E_{n,m}=(2n+|m|+1)\Omega \;.
\ee
Since they are not eigenstates of the transverse momentum operator $P^\perp$, the BLFQ basis elements mix states with the same intrinsic motion but with different transverse center-of-mass momenta. This is the price we pay for employing the 2D-HO states as single particle basis states in the transverse plane. We may employ, when needed, a Lagrange multiplier technique to enforce factorization of the transverse center-of-mass component of the amplitude from the internal motion components following techniques used in non-relativistic nuclear physics~\cite{Navratil:2000ww,Navratil:2000gs}.  We may also work with alternative coordinates chosen to achieve factorization~\cite{Maris_Cracow}.

The final quantum number, $\lambda$, labels the particle's helicity, which is the eigenvalue of $J^3_i$, see (\ref{two-part-decomp}). The electron (photon) helicity takes values $\lambda=\pm 1/2$ ($\lambda=\pm 1$).

We present only selected essentials of  our method; more details of the basis states may be found in Appendix~\ref{Transverse}. We note here that our transverse modes depend only on the combination $b:=\sqrt{M\Omega}$ (and not on $M$ and $\Omega$ individually). This is a free parameter which must be chosen. Since our goal is to design a basis which matches as closely as possible the symmetries of the QED Hamiltonian, we note that there is only one mass scale in QED, and that is the physical electron mass $m_e$. A sensible choice for our 2D-HO parameter is therefore\footnote{In Fock sectors with $n$ particles the effective 2D-HO parameter for the center-of-mass motion is $b_n^{cm}=\sqrt{nM\Omega}=b\sqrt{n}$, \ie, $\sqrt{n}$ times of that for single-particle states. Thus, in order to match the center-of-mass motion across different sectors as required by QED vertices, we adopt sector-dependent 2D-HO parameters $b_n=b/\sqrt{n}$ for Fock sectors with $n$ particles, where $b=m_e$ is the 2D-HO parameter in the one particle sector.} $b=m_e$, and we adopt this throughout.

Now, to see why this choice of basis is suited to light-front problems, we relate the single particle quantum numbers $\{k,n,m,\lambda\}$ to the segment numbers of the states $\alpha$. So, consider a multi-particle state $\ket{\alpha}=\otimes \ket{\bar\alpha}$, which belongs to a particular segment and is an eigenvector of $P^+$, $J^3$,  and $Q$ with eigenvalues $K$, $M_j$ and $N_f$, respectively. If $k^l$, $m^l$, $n^{f,l}$, and $\lambda^l$ are the quantum numbers for, respectively, the longitudinal momentum, longitudinal projection of angular momentum, net fermion number and helicity of the $l^\text{th}$ particle in the state then, summing over particles $l$, we have
\begin{align}
    \label{basis_symconstrain_k}
    &\sum_l k^{l}=K\;, \quad \sum_l n^{f,l}=N_f \; , \\
    &\sum_l m^{l}\equiv M_t \;, \quad \sum_l \lambda^l\equiv S \;, \\
    \label{basis_symconstrain_nf}
    &M_j = M_t+S \;.    
\end{align}
(The single particle net fermion number $n_f$ is 1 for $e$, -1 for $\bar e$ and 0 for $\gamma$.) We see that the basis states $\ket\alpha$ are eigenstates of $J^3_o$ and $J^3_i$ individually, with eigenvalues $M_t$ and $S$. Note, though, that it is the sum $M_j$ which is conserved by the light-front QED Hamiltonian.

While each basis state belongs to one and only one segment, it is clear that the basis states $\ket{\alpha}$ themselves are not eigenstates of QED (written as $\ket{\beta}$). These must still be constructed by diagonalizing $P^-_\text{QED}$ in this basis. For example, the physical electron eigenstate $\ket{\beta}{=}|e_\text{phys}\rangle$ can be expanded as 
\begin{align}\label{phys-expand}
    \ket{e_\text{phys}}=\sum_\alpha \ket{\alpha} \braket{\alpha}{e_\text{phys}} \;.
\end{align}
in which both the eigenstate on the left and the basis states on the right belong to the {\it same} segment. Diagonalizing the Hamiltonian in our basis would yield the coefficients $\braket{\alpha}{\beta}$, and hence the physical states $\ket{\beta}$. In order to do this, though we need to be able to implement our basis numerically, which requires some truncation. We turn to this now.

\subsection{Basis reduction} \label{3b}
%
Since a quantum field theory contains an infinite number of degrees of freedom, reduction of the basis space is necessary in order for numerical calculations to be feasible. For us, this reduction takes place both in the basis states retained (exploiting symmetries) and in the Fock space itself (\ie\ we retain only certain sectors and implement regulators).

The first type of reduction is called ``pruning'', in which we exclude basis states which are not needed for desired observables. The pruning process is lossless, in that it does not lead to loss of accuracy in the desired observables.  
For example, in bound state problems, one is typically interested in states with definite $N_f$ and $M_j$. 
Combining this with the longitudinal boost invariance inherent to light-front dynamics, one can choose $K$ based on the desired ``resolution'' for the longitudinal momentum partition among the basis particles~\cite{Brodsky:1997de}. Thus, one only needs to work in a single segment of the QED eigenspace, neglecting the others, without loss of information. From here on we write ``BLFQ basis'' to mean the basis of a {\it single} segment.

Pruning alone is not enough to reduce the basis space to finite dimension, however, since even a single segment contains an infinite number of degrees of freedom. To further reduce the basis dimensionality we need to perform basis {\it truncation}, which unavoidably causes loss of accuracy in calculating observables. Basis truncation is implemented at two levels.

{\it i) Fock-sector truncation.}
Consider the physical electron state. This has components in all Fock-sectors with $N_f=1$, which we write schematically as
\begin{align}
\label{state_expan_BLFQ_2}
|e_\text{phys}\rangle=a|e\rangle+b|e\gamma\rangle+c|e\gamma\gamma\rangle+d|ee\bar{e}\rangle+\ldots.
\end{align}
Included in this series are, for example, the bare electron $\ket{e}$ and its photon-cloud dressing, $\ket{e\gamma}$, $\ket{e\gamma\gamma}$ etc. Together, the bare fermion and its cloud of virtual particles comprise the observable, gauge invariant electron, as originally described by Dirac \cite{Dirac:1955uv,Lavelle:1995ty,Bagan:1999jf}. We implement basis truncation by assuming that higher Fock-sectors give (with an appropriate renormalization procedure implemented) decreasing contributions for the low-lying eigenstates in which we are mostly interested. (One motivation for this is the success of perturbation theory in QED). In this first paper, we make the simplest possible nontrivial truncation, which is to truncate our Fock-sectors to $\ket{e}$ and $\ket{e\gamma}$. Thus, in this truncated basis, the physical electron state would be given by only the first two terms of (\ref{state_expan_BLFQ_2}). This is enough to calculate physical wavefunctions accurate up to the first-order of the electromagnetic coupling $\alpha$. Due to its simplicity this Fock sector truncation has been typical of light-front Hamiltonian approaches such as Refs. 
\cite{Honkanen:2010rc,Chabysheva:2009ez}
though an extension to include the 2-photon sector has been successfully implemented in solving for the electron's
anomalous magnetic moment \cite{Chabysheva:2009vm}.

{\it ii) Truncation within Fock-sectors.} Fock-sector truncation is still not enough to reduce the basis to finite dimension; each Fock particle has an infinite number of (momentum) degrees of freedom. In BLFQ, truncations of the longitudinal and transverse degrees of freedom are realized separately, and differently.

Truncation of the longitudinal basis space is realized through the finite size of the $x^-$ direction. By imposing (anti--)~periodic boundary conditions, the longitudinal momentum $k$ for single particles can only take discrete values, see (\ref{longitudinal_momentum}). Therefore, in a given segment with total longitudinal momentum $K$, only a finite number of longitudinal momentum partitions is available for the particles in the basis states, since each particle's momentum must obey $0<k\le K$ and all the $k$'s must sum to $K$. For segments with larger $K$, more partitions of longitudinal momenta among particles are possible, allowing for a ``finer" description of the longitudinal degrees of freedom. Thus, $K$ also regulates the longitudinal degrees of freedom; bases with larger $K$ have simultaneously higher ultra-violet (UV) and lower infra-red (IR) cutoffs in the longitudinal direction.

Now consider the transverse part. Recalling from above that the transverse states are eigenstates of a 2D-HO, with energies (\ref{ENM}), we define the total transverse quantum number for multi-particle basis states $\ket{\alpha}$ as, 
\begin{align}
    \label{N_segment}
    N_\alpha=\sum_l 2n_l+| m_l |+1\;,
\end{align}
where the sum runs over all particles in the state. This number is used as the criterion for transverse basis truncation; all the retained basis states satisfy
\begin{align}
    \label{Nmax_truncation}
    N_\alpha \le N_\text{max} \;,
\end{align}
for some chosen $N_\text{max}$. Physically, this simply corresponds to restricting the total 2D-HO energy (summed over all particles). $N_\text{max}$ is specified globally across all Fock-sectors to ensure that the transverse motion in different Fock-sectors is truncated at the same energies. As shown in Appendix~\ref{CUTOFF}, $N_\text{max}$ determines both the UV and IR cutoffs for the transverse basis space, see also~\cite{Coon:2012ab,Furnstahl:2012qg}.

This brings us to the end of our discussion on the BLFQ basis itself, so let us summarize the approach so far. The eigenspace of QED breaks up into segments, labelled by $K$, $M_j$ and $N_f$. The BLFQ basis is a basis of states for such a segment, with each basis element carrying the same three quantum numbers as the segment itself. The basis elements themselves are collections of Fock particle states. For each Fock particle, 2D-HO states/plane-waves are employed to represent the transverse/longitudinal degrees of freedom. The Fock particle states carry four quantum numbers, $k$, $n$, $m$, $\lambda$, see above. A complete specification of a BLFQ basis requires 1) the segment numbers $K,M_j,N_f$, 2) the parameters $b$ and $L$ pertaining to the transverse oscillator basis and length of the longitudinal direction, respectively, and 3) two truncation parameters, namely the choice of which Fock sectors to retain, and the transverse truncation parameter $N_\text{max}$. (Recall, $K$ automatically serves as a longitudinal truncation parameter because $x^-$ is compact.) 

Such a basis is finite dimensional. It is then a straightforward matter to diagonalize the QED Hamiltonian in the BLFQ basis. This yields, as well as the eigenvalues of the Hamiltonian, a representation of the physical states of QED in terms of the BLFQ basis, as in (\ref{phys-expand}). 

We end this section with a few words on renormalization. Our focus in this paper is on an initial exploration of tBLFQ, and we neglect the necessary counter-terms when writing down our Hamiltonians. (Hence, we adopt physical values for the electron mass and charge.) Renormalization within the BLFQ framework is possible, via a sector-dependent scheme~\cite{Karmanov:2008br,Karmanov:2012aj,Zhao:2013xx,Chabysheva:2009ez}. For an application, see~\cite{Zhao:2013xx}, in which the scheme is implemented for the QED Hamiltonian; the resulting electron anomalous magnetic moment agrees with the Schwinger value to within 1\%.

\section{Time-dependent Basis Light-front Quantization (\lowercase{t}BLFQ)}
\label{sec:tBLFQ}
Now that we have the physical states of QED, we turn to the transitions between them as caused by an external field. Our Hamiltonian now consists of two terms,
\begin{align}
\label{hami_simple}	
	&P^-=P^-_\text{QED}+V \;,
\end{align}
in which the new term $V$ comprises the interactions introduced by the background (laser field), just as in (\ref{H_interact}). See  Appendix~\ref{hami_full} for the explicit form of the new interactions.

As discussed above, only a single segment of states is needed to address bound-state problems. The presence of background field terms $V$ means, in general, that the full Hamiltonian $P^-$ will not posses the symmetries associated with conservation of longitudinal momentum ($K$) and longitudinal projection of total angular momentum ($M_j$). ({\it Net} fermion number is not affected, of course.) In other words, the background field can cause transitions between QED eigenstates in different segments. In order to account for this, the BLFQ basis must be extended to cover several segments. We refer to a collection of multiple BLFQ basis segments with different $K$'s and $M_j$'s  as the ``extended BLFQ basis".

In fact, since our particular choice of background field~(\ref{laser-profile}) only adds longitudinal momentum (and lightfront energy) to the system, the transverse degrees of freedom remain untouched, and the symmetry associated with $J^3$ holds even with the laser field switched on. Therefore, for our current example, we only need to include segments with different total longitudinal momenta $K$.

\begin{figure*}[t]
\begin{center}
\raisebox{210pt}{\begin{Large}$\ket{\alpha}=$\end{Large}}\includegraphics[width=0.7\textwidth]{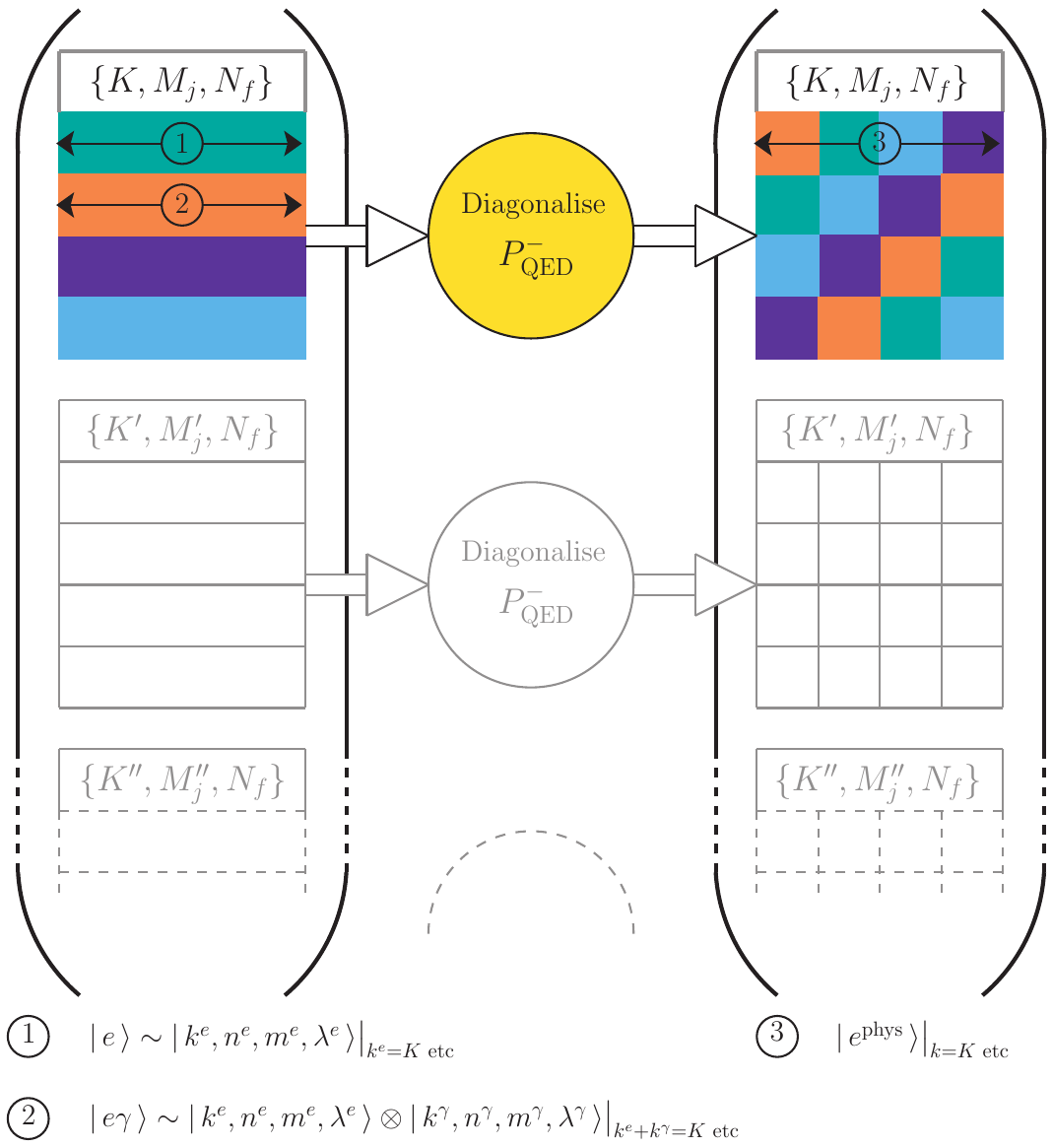}\raisebox{210pt}{\begin{Large}$=\ket{\beta}$\end{Large}}
\caption{\label{BASES} The BLFQ and tBLFQ bases. On the left, the extended BLFQ basis $\ket\alpha$. This is a collection of bases in different segments, each segment labelled by $K$, $M_j$ and $N_f$. (Since nothing in our theory changes net fermion number, all segments of interest have fixed $N_f=1$, in our case.) The states in each segment are bare states. Two such states, a bare electron and a bare electron + a photon, are illustrated. The BLFQ procedure diagonalizes the Hamiltonian in each segment. The basis states in $\ket{\alpha}$ are then rearranged into eigenstates $\ket{\beta}$ of the QED Hamiltonian, shown on the right. These are the tBLFQ basis states. }
\end{center}
\end{figure*}

We therefore begin by applying BLFQ to $P^-_\text{QED}$ in each segment, finding the physical states in that segment and representing them as in (\ref{phys-expand}). The combination of all such eigenstates from all the segments forms the ``tBLFQ basis'', which is a basis of physical eigenstates of QED. From here on we will write $\ket\alpha$ to represent the extended BLFQ basis, and $\ket\beta$ to represent the tBLFQ basis of physical states. See Fig.~\ref{BASES} for an illustration of the two different bases and the relationship between them.

In constructing the tBLFQ basis, one needs to specify the number of the segments to be included. Larger, less-truncated, basis spaces yields more realistic and detailed descriptions of the underlying system. The price we pay for increased basis dimensionality is of course increased computational time.

We have reached the stage at which we have an (appropriate) set of physical eigenstates of QED. We now describe the preparation of the initial state, and its evolution in time under the Hamiltonian (\ref{hami_simple}).

\subsection{Initial state preparation}
In perturbation theory, scattering calculations take initial states to be eigenstates of the {\it free} part of the Hamiltonian, following the usual assumption of asymptotic switching, see though~\cite{Kulish:1970ut,Horan:1999ba}. In our calculations, initial states are taken to be physical eigenstates of QED. For our nCs, process, for example, the initial state is a single {\it physical} electron with longitudinal momentum $K_i$. This state can be identified as the ``ground state'' of the QED Hamiltonian $P^-_\text{QED}$ in the segment $N_f{=}1$, $M_j{=}\tfrac{1}{2}$ and $K$=$K_i$ (since there is no other state in that segment with a lower energy). In the tBLFQ basis, which is just the set of eigenvectors of QED, this initial state is trivially defined.

\subsection{State evolution}
Recalling the discussion in Section~\ref{sec:bg}, our initial state evolves, in the interaction picture, according to
\begin{align}
\label{sol_wave-eq_i}
	\ket{\psi; x^+}_I &= \mathcal{T}_+ e^{-\frac{i}{2}\int\limits_0^{x^+} V_I}\ket{\psi;0}_I \;,
\end{align}
in which $\ket{\psi;0}_I$ is the initial state, equal to a chosen eigenstate of QED (or a superposition thereof). In general, the interaction operator $V_I$ will not commute with itself at different times. We decompose the time-evolution operator into many small steps in light-front time $x^+$, introducing the step size $\delta x^+$,
\begin{align}
\label{sol_wave-eq_i_discrete}
\mathcal{T}_+ e^{-\frac{i}{2}\int\limits_0^{x^+} V_I} \rightarrow &\big[1-\tfrac{i}{2}V_I(x^+_{n})\delta x^+\big] \cdots \big[1-\tfrac{i}{2}V_I(x^+_{1})\delta x^+\big] \;,
\end{align}
in which each square bracketed term is a matrix, and we let each of these matrices act on the initial state sequentially. Between each matrix multiplication we insert a (numerically truncated) resolution of the identity, so that the evaluation of (\ref{sol_wave-eq_i_discrete}) amounts to the repeated computation of the overlaps
\begin{align}
\label{VLAS_trans_interact}
	\bra{{\beta'}} V_{I} \ket{\beta} &= \bra{{\beta'}}V\ket{\beta}  \exp{[\tfrac{i}{2}(P^-_{\beta'}-P^-_{\beta})x^+]},
\end{align}
in which the $P^-_\beta$ are the previously solved eigenergies of $P^-_\text{QED}$, and their presence follows from Eq.~(\ref{Energi}). In order to calculate the left hand side of (\ref{VLAS_trans_interact}) in our numerical scheme, it is simpler to first calculate the phase factor and then calculate the remaining overlap in terms of the (extended) BLFQ basis, as follows:
\begin{align} \label{VLAS_trans}
	\bra{{\beta'}}V\ket{\beta}=\sum_{\alpha'\alpha} \braket{{\beta'}}{{\alpha'}} \bra{{\alpha'}}V\ket{{\alpha}} \braket{{\alpha}}{\beta} \;.
\end{align}
The resulting interaction picture matrix elements are the elementary building-blocks for evaluating all observables. 

For our particular choice of background field, the structure of the matrix elements between BLFQ basis elements $\ket{\alpha}$ is simple. The interaction terms introduced by the chosen background do not contain the quantum gauge field (see Appendices~\ref{hami_full} and~\ref{laserme_comp} for details), and therefore do not {\it directly} connect different Fock sectors; matrix elements of the type $\bra{\alpha'(e)}V\ket{\alpha(e\gamma)}$ are therefore all zero. (Physically, the only direct effect of the chosen background field is to either increase or decrease the longitudinal momentum $k$ of an electron by $k_\text{las}$.) Matrix elements between the same Fock sectors (in our case $\bra{\alpha'(e)}V\ket{\alpha(e)}$ and $\bra{\alpha'(e\gamma)}V\ket{\alpha(e\gamma)}$), on the other hand, are nonzero. If $\bar\alpha$ and $\bar\alpha'$ label two Fock electron states, then one finds for example
\be
	\bra{\bar{\alpha}'}V\ket{\bar\alpha} = m_e a_0 \big(\delta_{\lambda}^{\lambda'} \delta_{n}^{n'} \delta_{m}^{m'}\big)\ (\delta_{k}^{k'+k_\text{las}}+\delta_{k}^{k'-k_\text{las}}) \;,
\ee
in which $\delta^*_*$ is the Kronecker delta. The magnitude of the matrix element is proportional to the field intensity $a_0$. It is the sum of two terms, originating in the two exponentials in (\ref{laser-profile}). Each term is the product of two Kronecker deltas. The first delta conserves all quantum numbers between the states except for the longitudinal momentum (since that is all that our background field alters). The second delta fixes the difference between the $k$ values of the basis elements to be $k'=k\pm k_\text{laser}$; this is simply the ``conservation'' of longitudinal momentum among the initial and final electrons, in that any added energy-momentum must come from the laser field.

\subsection{Numerical Scheme}\label{NumericalScheme}
A direct implementation of Eq.~(\ref{sol_wave-eq_i_discrete}) leads to the so-called Euler scheme which relates the state at $x^+{+}\delta x^+$ to that at $x^+$; this  scheme is however not numerically stable (since it is not symmetric in time) and the norm of the state vector $\ket{\psi;x^+}$ increases as time evolves, see Ref.~\cite{Iitaka:1994}. We therefore adopt the second order difference scheme MSD2~\cite{Askar:1978}, which is a symmetrized version of the Euler scheme relating the state at $x^+{+}\delta x^+$ to those at $x^+$ and $x^+{-}\delta x^+$ via
\begin{align}
    \label{evolve_MSD2_scheme}
    \nonumber &\ket{\psi;x^+{+}\delta x^+}_I \\
    &=\ket{\psi;x^+{-}\delta x^+}_I +(e^{-iV_I\delta x^+/2}-e^{iV_I\delta x^+/2})\ket{\psi;x^+}_I\nonumber \\
    &\approx \ket{\psi;x^+{-}\delta x^+}_I -iV_I(x^+)\delta x^+\ket{\psi;x^+}_I \;.
\end{align}
It can be shown that the MSD2 scheme is stable, with the norm of the states conserved, provided that $|V_{I;\text{max}}|\delta x^+<1$, where $V_{I;\text{max}}$ is the largest (by magnitude) eigenvalue of $V_I$~\cite{Iitaka:1994}. This requirement imposes an upper limit on the step size $\delta x^+$. Further limits on $\delta x^+$ will be discussed below.

(Note that in order to provide sufficient initial conditions for the MSD2 scheme, we use the standard Euler scheme to evolve the initial state one half-step forward, generating $\ket{\psi;\delta x^+/2}_I$. Then we use the MSD2 scheme to evolve $\ket{\psi;\delta x^+/2}_I$ an additional half-step forward, generating $\ket{\psi;\delta x^+}_I$. With both $\ket{\psi;0}_I$ and $\ket{\psi;\delta x^+}_I$ available the MSD2 scheme is ready to generate $\ket{\psi;x^+}$ at subsequent times, in time steps of $\delta x^+$.)

This concludes our discussion of the principles behind, and the method of application, of BLFQ and tBLFQ. The reader interested in more details is referred to Appendix~\ref{OpBLFQbasis_BLFQ_basics} for the (analytic) representation of states and operators in the BLFQ basis, and to Appendix~\ref{sample_BLFQ} for a worked example of the construction of a small, simple BLFQ basis, diagonalization of the Hamiltonian and an example tBLFQ calculation.

In the next section we turn to the results of our calculation of the nCs process. 

\section{Numerical Results}
\label{sec:result}
In this section we present numerical results for non-linear Compton scattering (nCs), computed in the tBLFQ framework. Since the laser matrix elements $\bra{\beta'}V\ket{\beta}$ play an important role in the numerical results, we first check them against those from light-front perturbation theory, in Section~\ref{ssec:cali}. We then perform a systematic study of nCs using the laser matrix elements obtained from BLFQ, in Section~\ref{ssec:nonpert}. For interested readers we present the full details in the numerical calculation for the nCs process (in a ``minimal'' basis) in Appendix.~\ref{sample_BLFQ}.
\subsection{Comparison of laser matrix elements}
\label{ssec:cali}
The laser matrix elements
\be\label{final}
	\bra{\beta'}V\ket{\beta}
\ee
 are calculated in the BLFQ framework from the wavefunctions of $\ket\beta$ and $\ket{\beta'}$ found from diagonalizing $P^-_\text{QED}$. Due to the small value of the electromagnetic coupling $\alpha=e^2/(4\pi)$ these wavefunctions can also be calculated in perturbation theory, and we will use this to check the BLFQ procedure.
  
Let us begin with the perturbative calculation of the matrix element (\ref{final}). The background field enters only as an operator sandwiched between the states. What we must do is to construct the QED eigenstates $\ket\beta$. This can be achieved using ordinary, time-independent perturbation theory. To be concrete we will take $\ket{\beta}=\ket{e_\text{phys}}$, the physical electron, and $\ket{\beta'}=\ket{e\gamma_\text{scat}}$, the electron-photon scattering state.  We will work to first order in the coupling. So, if $\ket{j}$ is a complete set of eigenstates of the {\it free} light-front Hamiltonian $P^-_\text{free}$, and the QED interaction linear in $e$ is $V_Q$ (see the first line of (\ref{FULL})), then the physical electron can be written, to first order, 
\be
	\ket{e_\text{phys}}=\ket{e}- \sum\limits_{j\not=e} \ket{j}\frac{\bra{j}V_Q\ket{e}}{P_\text{free}^\LCm(j) - P_\text{free}^\LCm(e)} \;.
\ee
Similarly, the physical electron-photon state is
\be
	\ket{e\gamma_\text{scat}}=\ket{e \gamma}  - \sum\limits_{j\not=e\gamma} \ket{j}\frac{\bra{j}V_Q\ket{e\gamma}}{P_\text{free}^\LCm(j) - P_\text{free}^\LCm(e\gamma)} \;.
\ee
The matrix element (\ref{final}) is therefore approximated in perturbation theory by 
\be\begin{split}\label{pert-res}
	\bra{e\gamma_\text{scat}} V \ket{e_\text{phys}} &= \bra{e\gamma} V \frac{1}{P^\LCm_\text{free}(e)-\hat{P}^\LCm_\text{free}}V_Q\ket{e} \\
	& +\bra{e\gamma} V_Q \frac{1}{P^\LCm_\text{free}(e\gamma)-\hat{P}^\LCm_\text{free}}V\ket{e} \;,
\end{split}
\ee
in which we have written a hat over the operator $P^\LCm_\text{free}$ to distinguish it from the eigenvalues $P^\LCm_\text{free}(e)$ and $P^\LCm_\text{free}(e\gamma)$. Note that $\ket{e}$ and $\ket{e\gamma}$ are eigenstates of $P^-_\text{free}$, but in order to compare with the BLFQ calculation we need to evaluate the matrix elements (\ref{pert-res}), and hence the states, in the BLFQ basis $\ket{\alpha}$. The calculation is uninstructive, so we simply present the result in  Appendix~\ref{laserme_comp}. 

Now, how do we compare this with a BLFQ calculation? We begin by constructing a basis containing only two ($K$-)segments, $K=K_i$ and $K=K_i+k_\text{las}$, using the same parameters as in the perturbative calculation. In the $K=K_i$ segment we retain only the single electron (ground) state; this acts as the initial state. In the $K=K_i+k_\text{las}$ segment we retain only the electron-photon (excited) states. Such a basis, while heavily truncated, is all that is required for comparing the results of BLFQ and tBLFQ with the perturbative result (\ref{pert-res}).
 
In general we would expect that the two matrix elements will match in the case of small QED coupling $\alpha$. However, in our truncated Fock-space (with only the one-electron and one-photon-one-electron sectors) there is one further source of potential discrepancy, for the following reason: the perturbative matrix elements (\ref{pert-res}) are calculated in the (complete) momentum basis first and then projected onto the initial ($\ket{e}$) and final ($\ket{e\gamma}$) states in the BLFQ basis. The BLFQ matrix elements, on the other hand, are calculated in a truncated basis space throughout. In the language of perturbation theory, there exists extra truncation effects in the BLFQ matrix elements between the ``propagator'' and the QED vertices, cf. Eq.~(\ref{pert-res}). Due to this ``intermediate" basis truncation, exact agreement can only be expected in the continuum limit ($N_{\rm{max}}\to\infty$).

Direct comparison as a function of $N_{\rm{max}}$ is difficult, because as $N_{\rm{max}}$ increases the spectrum of $P^-_\text{QED}$ changes (more states appear in the spectrum) and it becomes difficult to keep track of the $N_{\rm{max}}$ dependence for specific matrix elements. We will now look at a test case, the nCs process with perturbative and nonperturbative matrix elements as inputs, and compare their predictions for the population of various tBLFQ basis states as a function of $N_{\rm{max}}$.

We take $K_i{=}1.5$, $N_\text{max}{=}16$, and parameters $a_0=10$, $k_\text{las}{=}2$ for the laser profile~(\ref{laser-profile}). Thus the laser can cause transitions between the $K{=}1.5$ and $K{=}3.5$ segments. In general, we expect the truncation error between perturbative and nonperburbative matrix elements to diminish as $N_\text{max}$ increases and more basis states are used. However, since we are neglecting various renormalization counter-terms, at sufficiently large $N_\text{max}$, high order and divergent loop effects may lead to further discrepancies between the (leading order) perturbative and nonperburbative matrix elements. Contributions from these loop effects are proportional to (higher-than-leading) powers of $\alpha$ and generally increase with the ultraviolet and/or infrared cutoff $N_\text{max}$. Since at this stage we are interested only in verifying the decreasing truncation error as $N_\text{max}$ increases, we suppress here the contribution of loop effects by artificially reducing the coupling constant so that $e^2/(4\pi)\to$1/13700.
\begin{figure}[!t]
\centering
\includegraphics[width=0.47\textwidth]{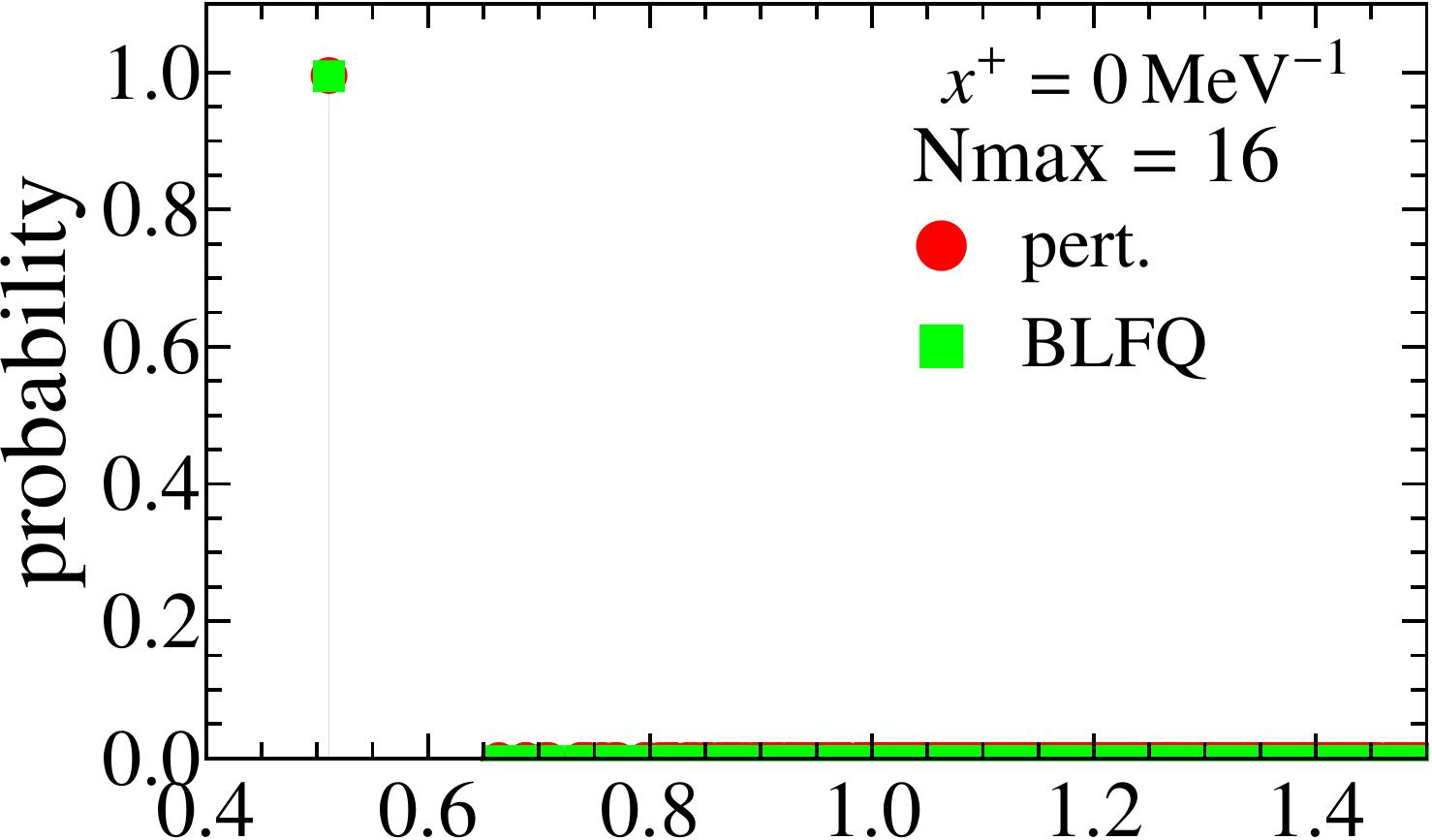}
\includegraphics[width=0.47\textwidth]{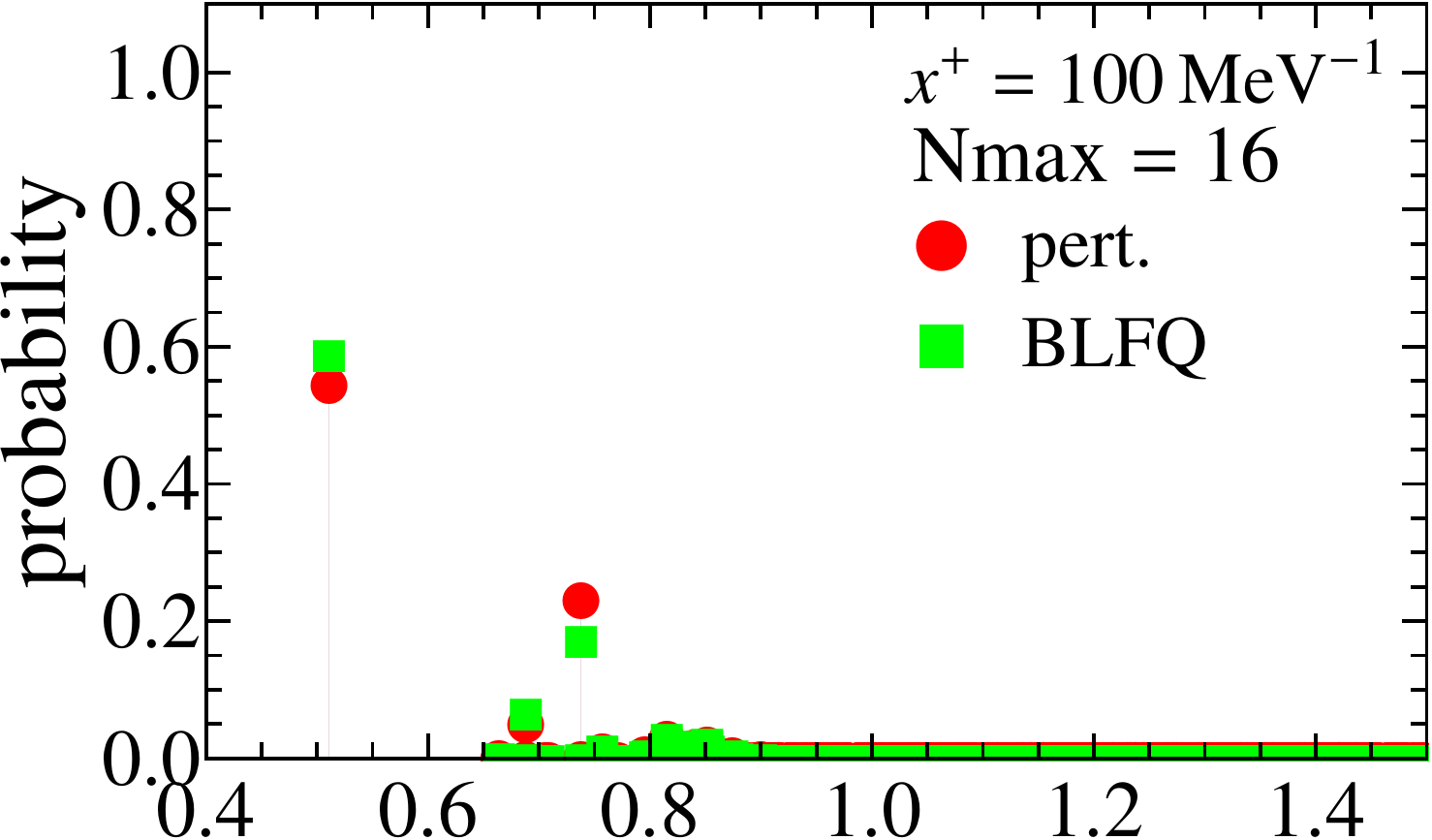}
\includegraphics[width=0.47\textwidth]{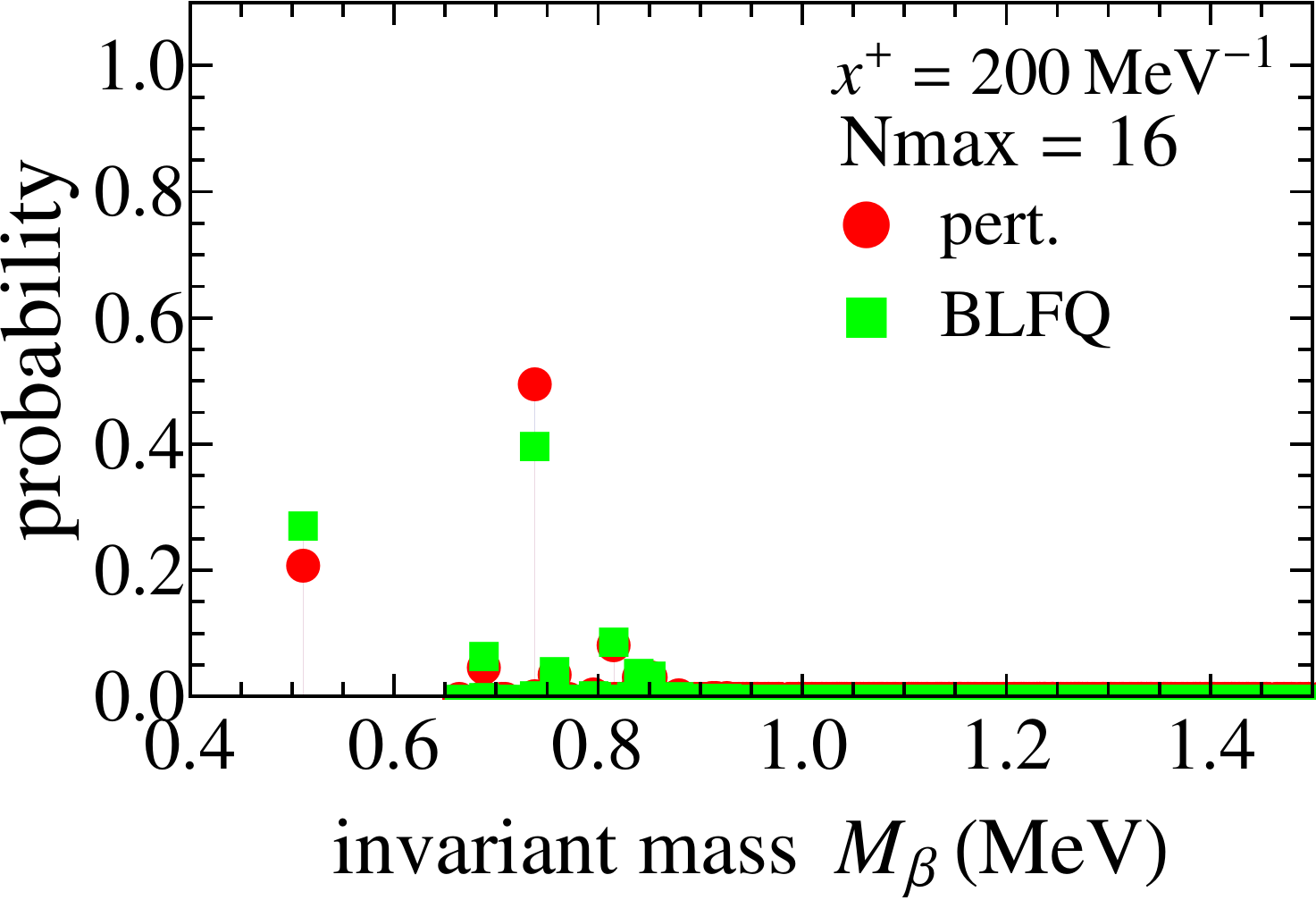}
\caption{\label{fig:state_evol} (Color online) Time evolution of the electron system in the laser field (at $N_\text{max}$=16). Upper, middle and lower panels correspond to exposure time $x^+$=0, 100, 200 MeV$^{-1}$ respectively (the laser field is switched on at $x^+$=0). Each dot on these plots corresponds to a tBLFQ basis state $\ket{\beta}$ in $K$=3.5 segment. Y-axis is the probability, $|c_\beta|^2$, for each basis state and x-axis is the corresponding invariant mass, $M_\beta$. Green (red) dots are results based on laser matrix elements evaluated nonperturbatively (perturbatively). Note that the electromagnetic coupling constant $\alpha{=}e^2/(4\pi)$ is reduced to 1/13700, see text for details.
}
\end{figure}

\begin{figure}[t!]
\centering
\includegraphics[width=0.46\textwidth]{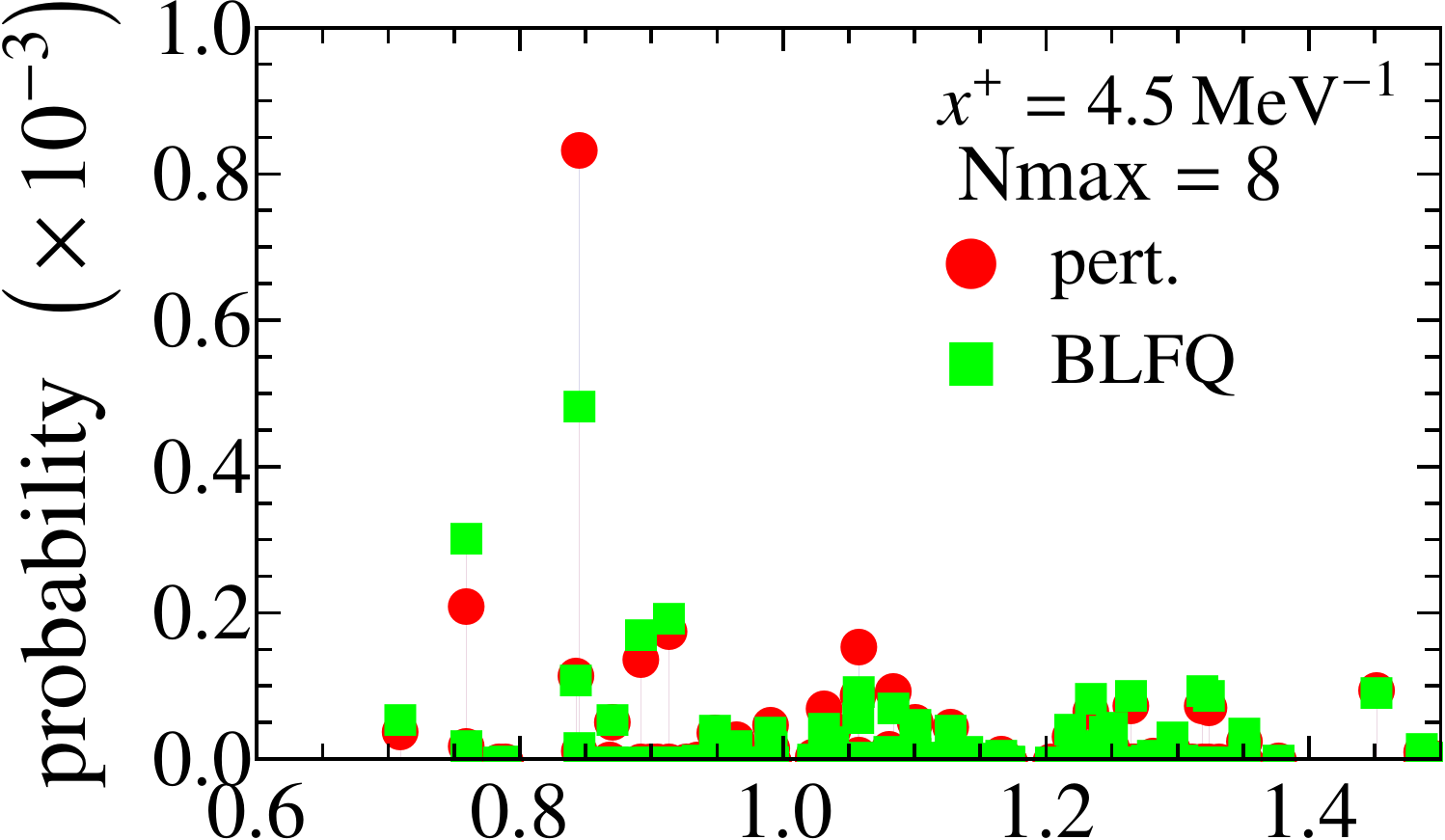}
\includegraphics[width=0.46\textwidth]{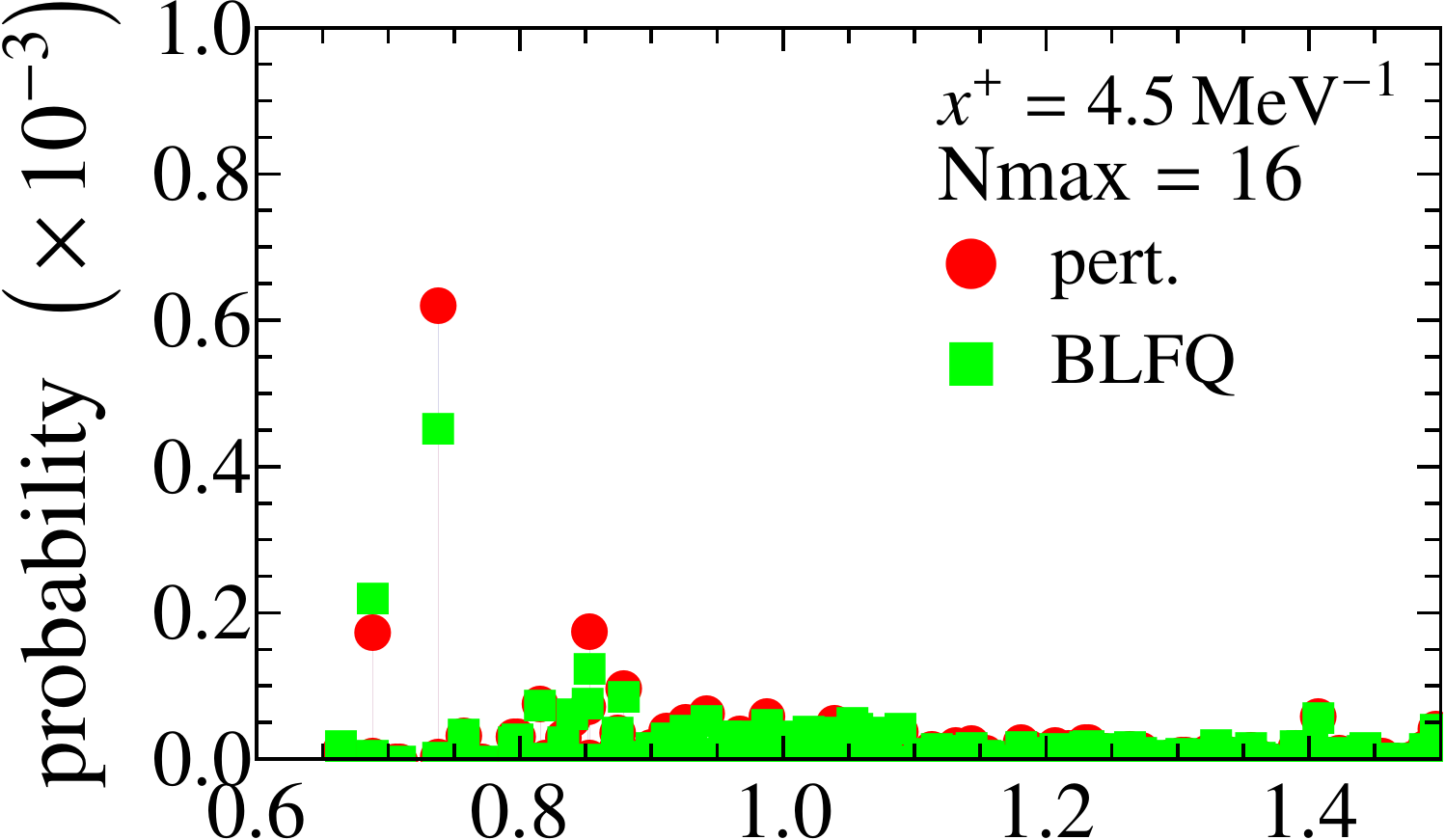}
\includegraphics[width=0.46\textwidth]{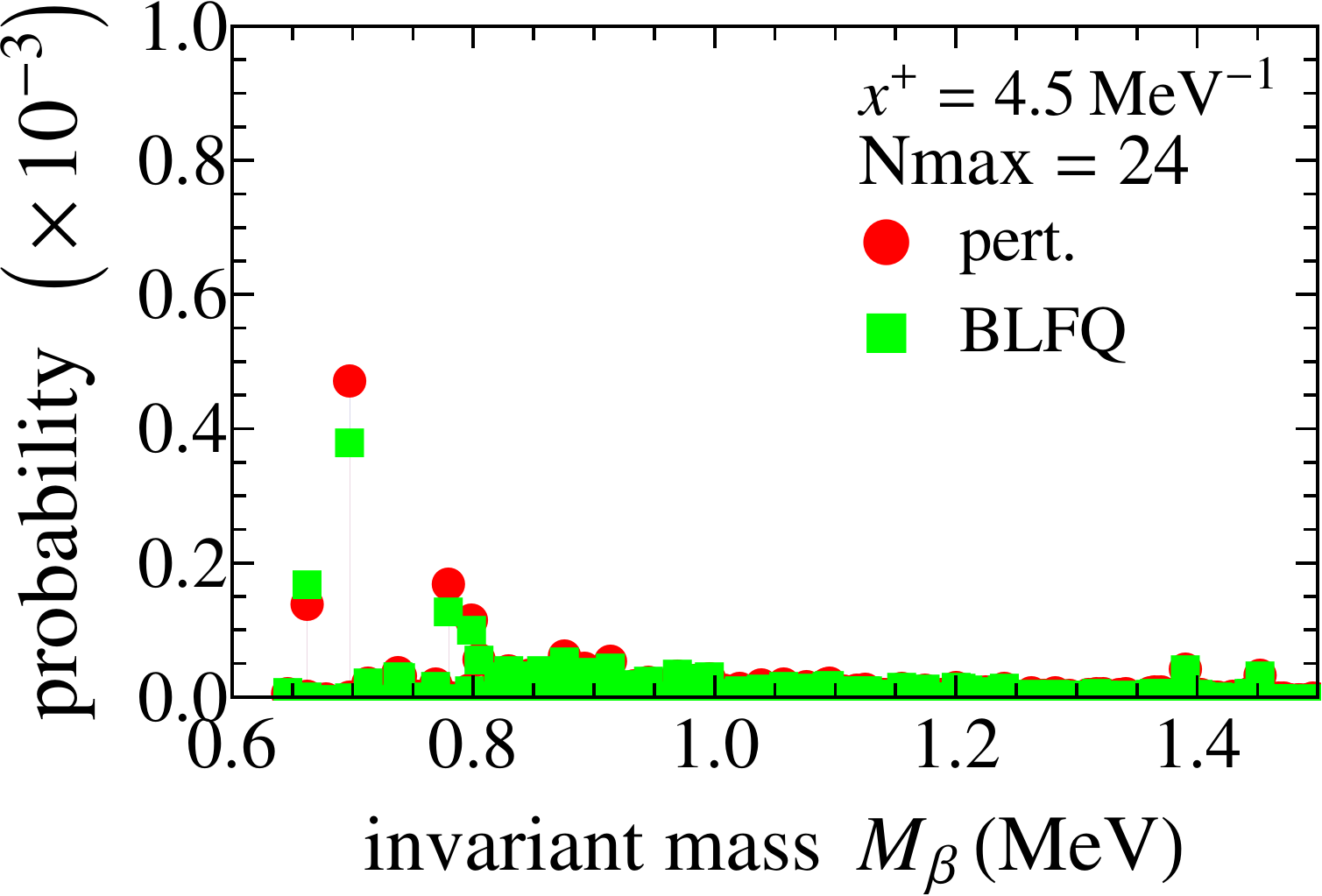}
\caption{\label{fig:state_evol2}  (Color online) ``Snapshots'' of the system at $x^+=4.5$ MeV$^{-1}$ in bases of $N_{\rm max}$=8 (upper panel), 16 (middle panel) and 24 (lower panel). Each dot on these plots corresponds to a tBLFQ basis state $\ket{\beta}$ in $K$=3.5 segment. Y-axis is the probability, $|c_\beta|^2$, (on a greatly expanded scale compared to Fig.~\ref{fig:state_evol}) for each basis state and x-axis is the corresponding invariant mass, $M_\beta$. Green (red) dots are results based on laser matrix elements evaluated nonperturbatively (perturbatively). Note that the electromagnetic coupling constant $\alpha{=}e^2/(4\pi)$ is reduced to 1/13700, see text for details.
}
\end{figure}
\begin{figure}[th!]
\centering
\includegraphics[width=0.48\textwidth]{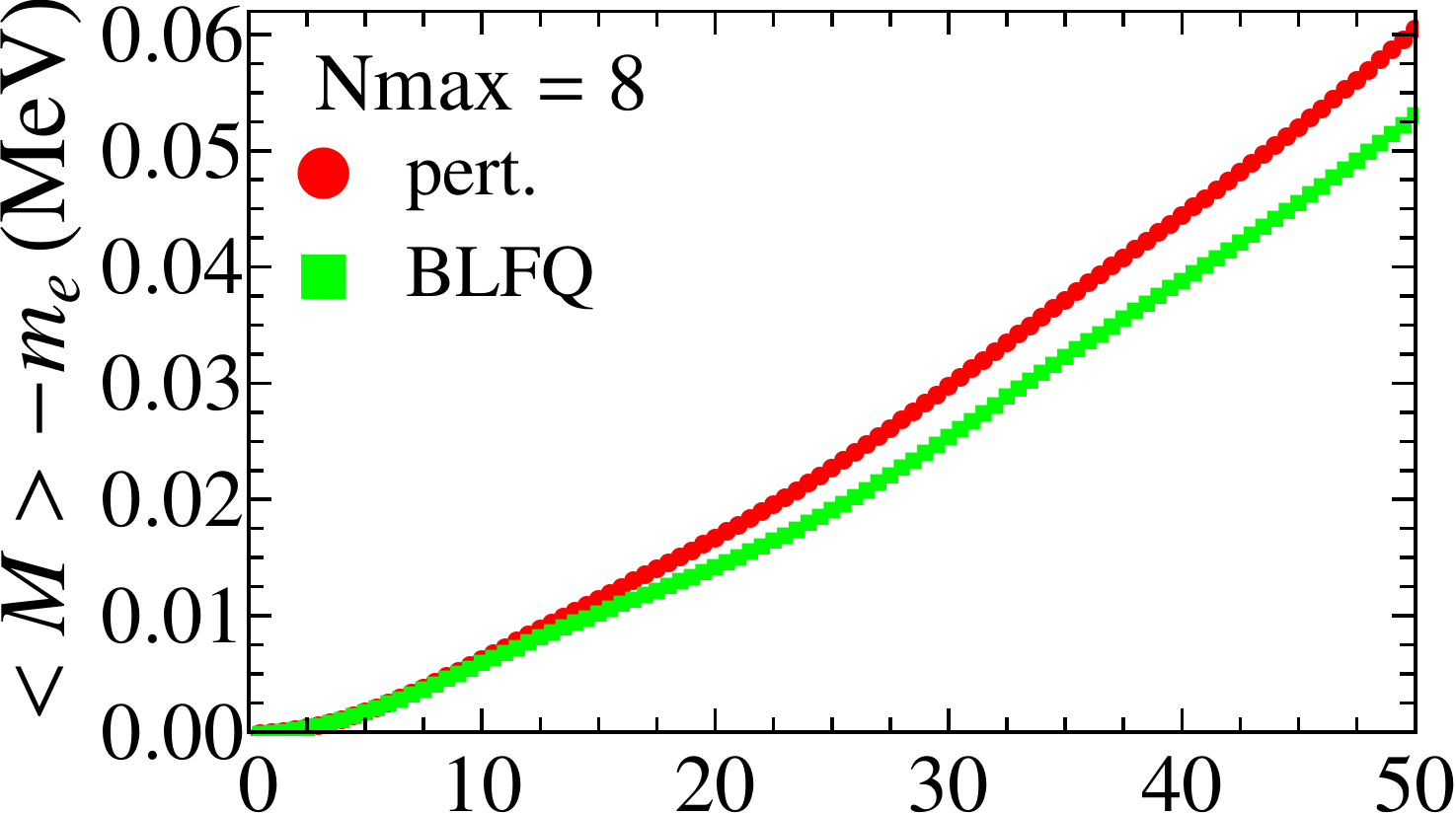}
\includegraphics[width=0.48\textwidth]{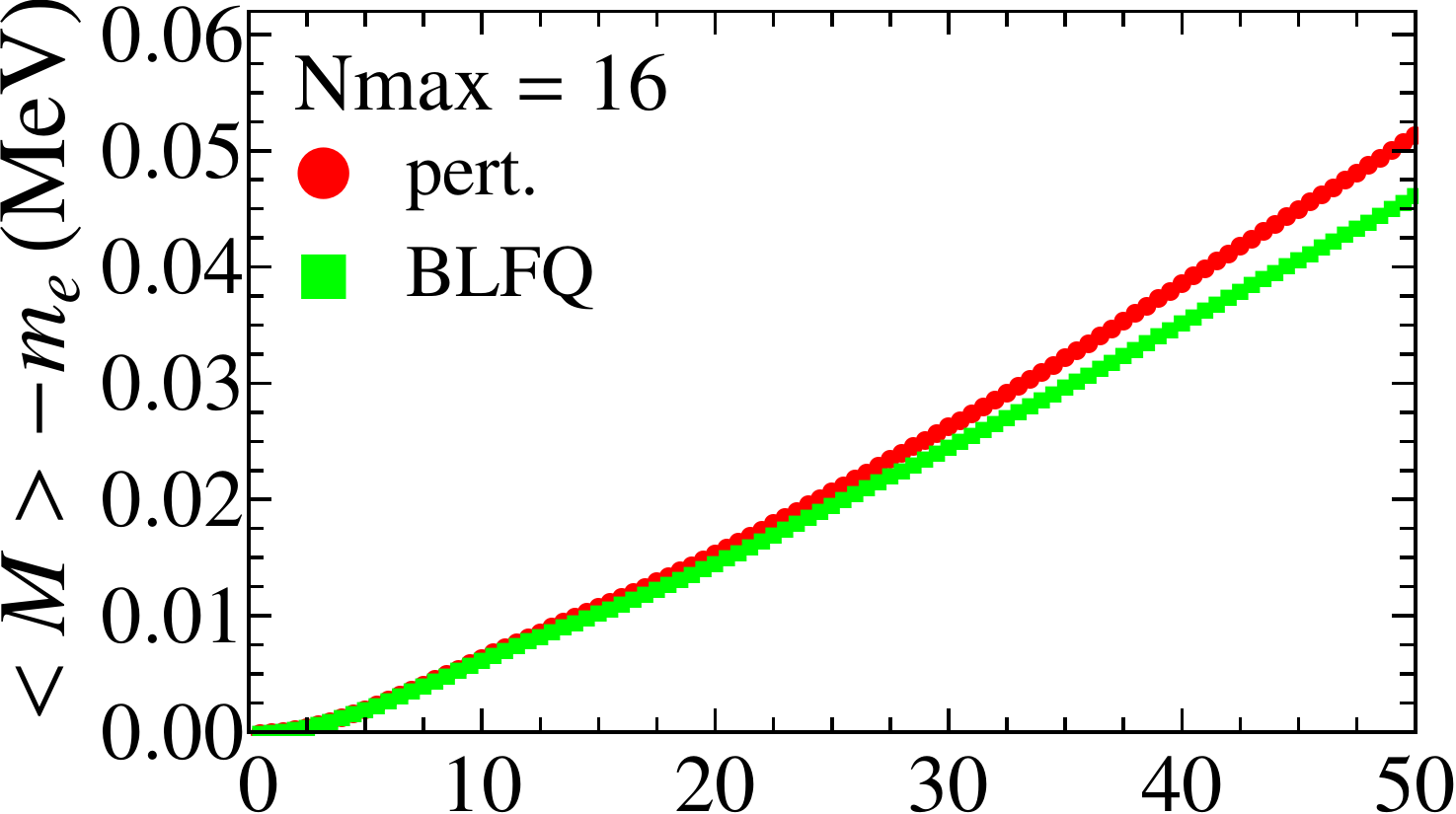}
\includegraphics[width=0.48\textwidth]{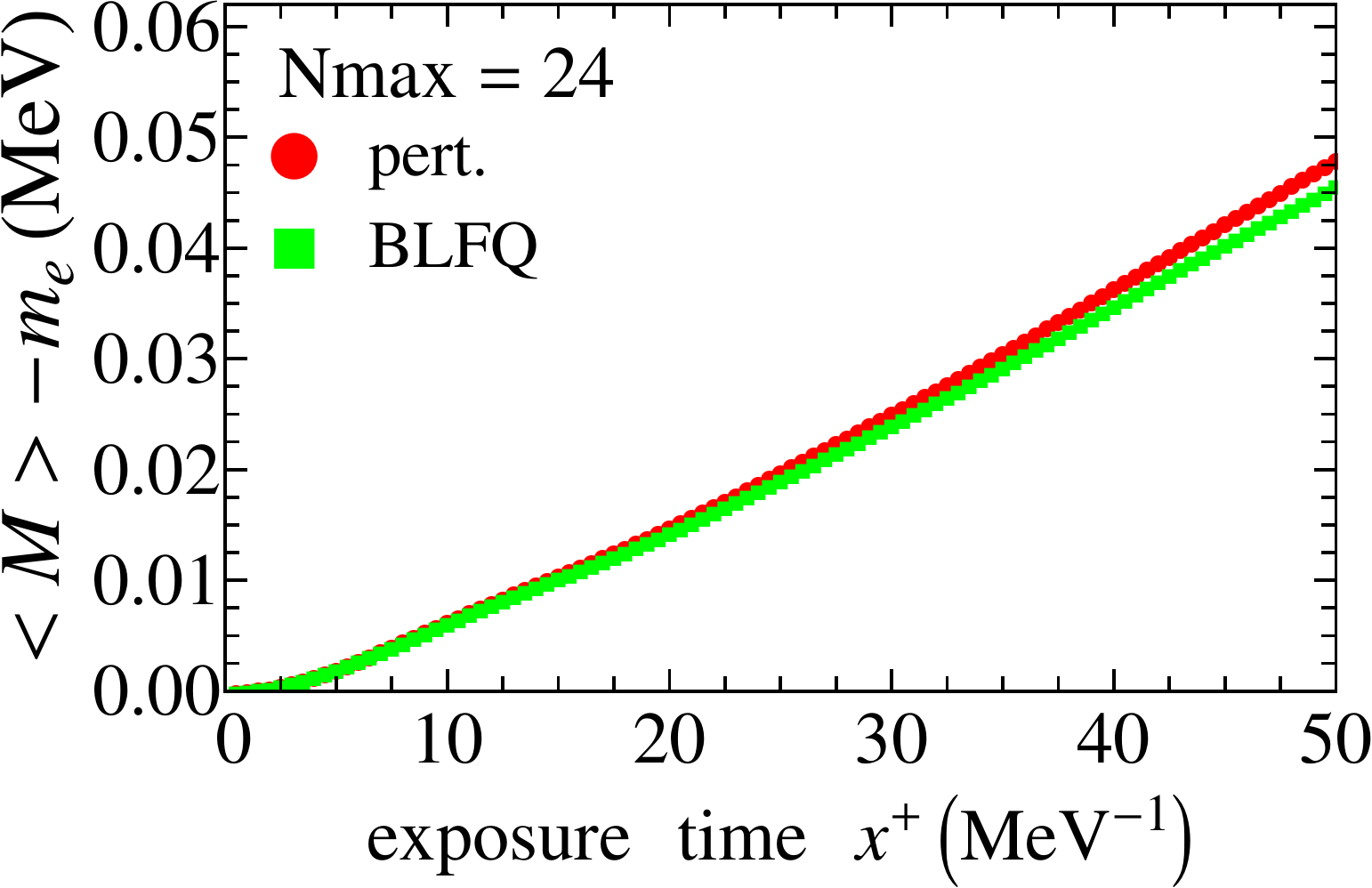}
\caption{\label{fig:invmass} (Color online) Time evolution of the average invariant mass of the electron system. Up, middle and lower panels are calculated in tBLFQ basis space with $N_\text{max}$=8, 16, 24 respectively. Y-axis is the difference between the average invariant mass $\langle M\rangle$ of the system at $x^+$ and that of a single electron $m_e$. X-axis is the (lightfront) exposure time $x^+$. Green (red) dots are results based on (non)perturbative laser matrix elements. Note that the electromagnetic coupling constant $\alpha{=}e^2/(4\pi)$ is reduced to 1/13700, see text for details.}
\end{figure}

At $x^+$=0 we switch on the laser field and evolve the initial single electron state according to Eq.~(\ref{evolve_MSD2_scheme}). The population of tBLFQ basis states (the probabilities $|c_\beta(x^+)|^2$) in the $K{=}3.5$ segment, as a function of light-front time, are shown in Fig.~\ref{fig:state_evol}, along with the corresponding perturbative results. Different tBLFQ basis states are distinguished by their respective invariant masses, $M_\beta$, as defined in Eq.~(\ref{inv_mass}).
As time evolves the probability for the single electron state (with invariant mass $\sim$0.511MeV) drops and various electron-photon states (with invariant mass above 0.6MeV) in the $K$=3.5 segment are gradually populated. At $x^+{=}100$ and $200$~MeV$^{-1}$, a peak structure is seen around the invariant mass of 0.74~MeV.  This can be understood as follows: because our laser profile~(\ref{laser-profile}) is only trivially dependent on light-front time, in that it switches on and off but is otherwise constant, in the infinite time limit only transitions between basis states with the same light-front energy can accumulate (the transition amplitudes between states with unequal energies oscillate with a period inversely proportional to their energy difference).  In this case the light-front energy of the initial (single electron) state is $P^-{=}m^2_e/K_i{=}0.17$~MeV, basis states in the $K{=}3.5$ segment with invariant mass around $M_{pk}{=}\sqrt{P^-K}{=}\sqrt{0.17\times3.5}{=}0.78$~MeV will thus accumulate and form a peak. The full peak develops over longer times and is located at approximately 0.8~MeV, independent of $N_\text{max}$.

In order to study the convergence between the perturbative and nonperturbative laser matrix elements as a function of $N_\text{max}$, we consider snapshots of the system at a fixed exposure time ($x^+{=}4.5$~MeV$^{-1}$) calculated in bases with increasing $N_\text{max}$ in Fig.~\ref{fig:state_evol2}. As expected the overall agreement between the results from perturbative and nonperturbative laser matrix elements indeed improves systematically as the basis $N_\text{max}$ increases.

As a measure of the energy transfer between the system and the laser field, we calculate the evolution of the average invariant mass $\langle M(x^+)\rangle{=}\sum{M_\beta|c_{\beta}(x^+)|^2}$ of the system as a function of the exposure time $x^+$. The numerical results calculated in bases with $N_\text{max}$=8,16,24 are compared in Fig.~\ref{fig:invmass}.

As the exposure time increases, the laser field pumps energy into the system, and the invariant mass of the system increases accordingly, as seen from Fig.~\ref{fig:invmass}. Again, as expected the agreement between the results from perturbative and nonperturbative laser matrix elements improves as the basis size ($N_{\rm max}$) increases. 

In the next subsection we will study the nCs process systematically in a larger basis space using laser matrix elements from the BLFQ approach.

\subsection{Numerical results for nCs}
\label{ssec:nonpert}

\begin{figure*}[!t]
\centering
\includegraphics[width=0.45\textwidth]{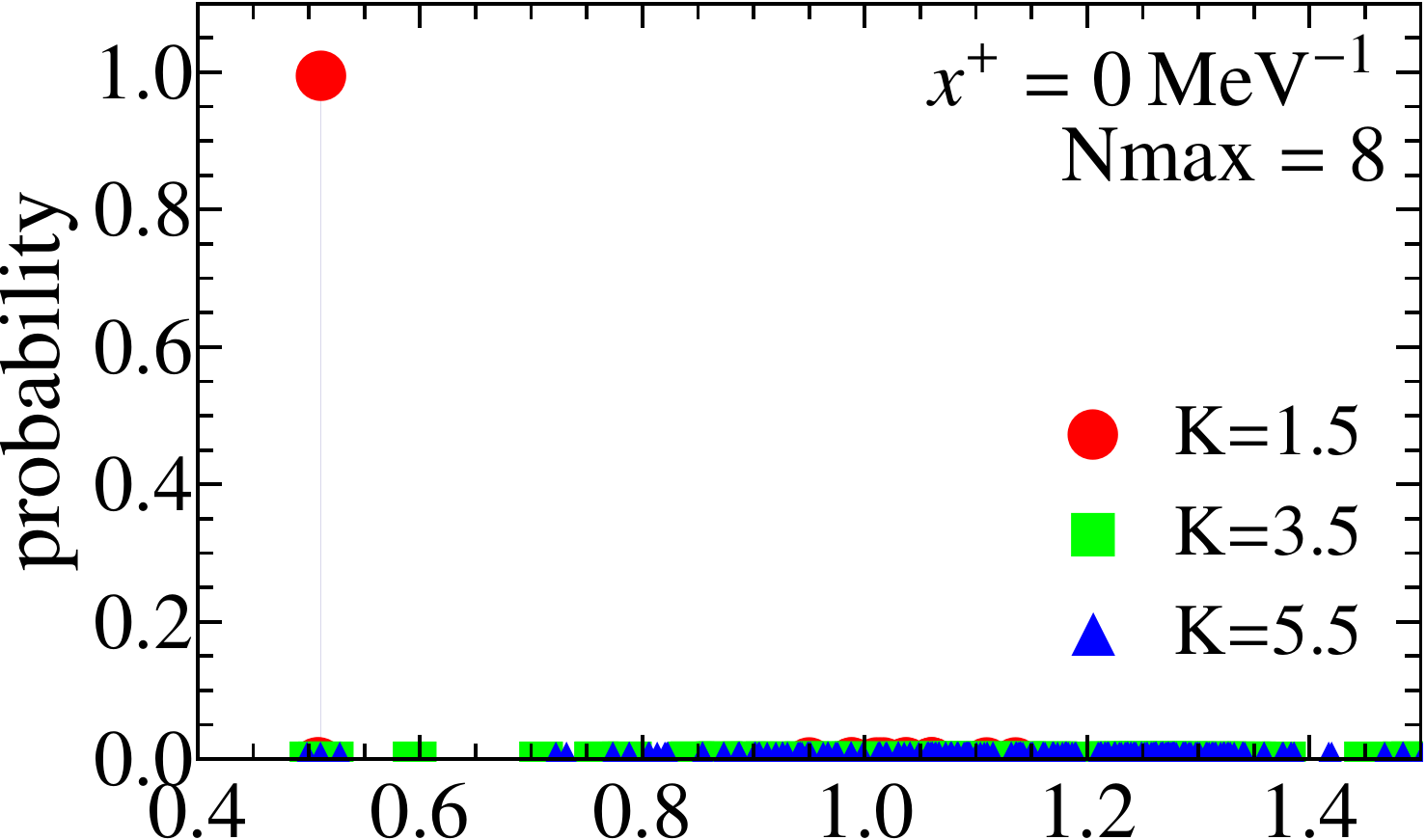}
\includegraphics[width=0.46\textwidth]{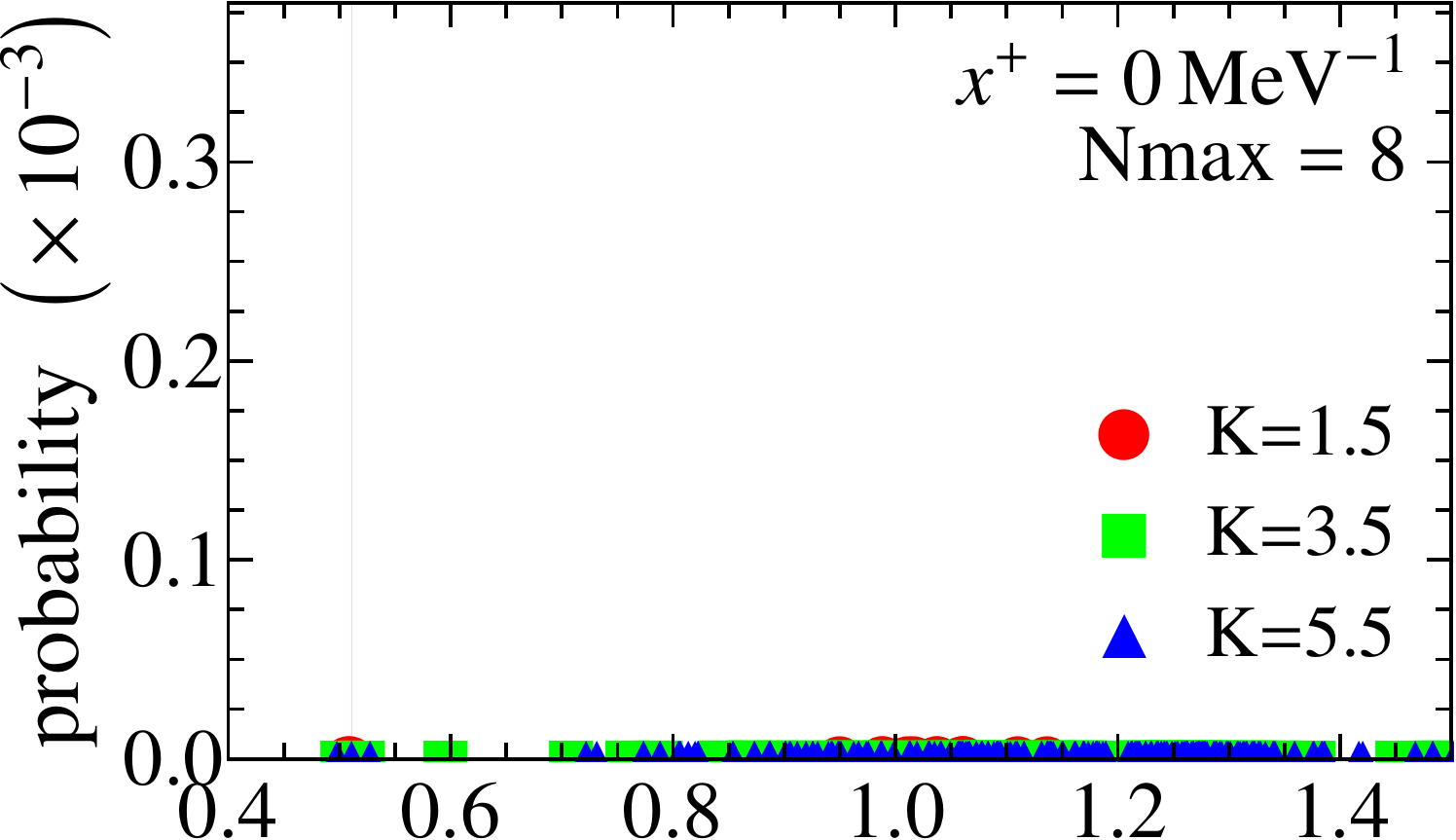}
\includegraphics[width=0.45\textwidth]{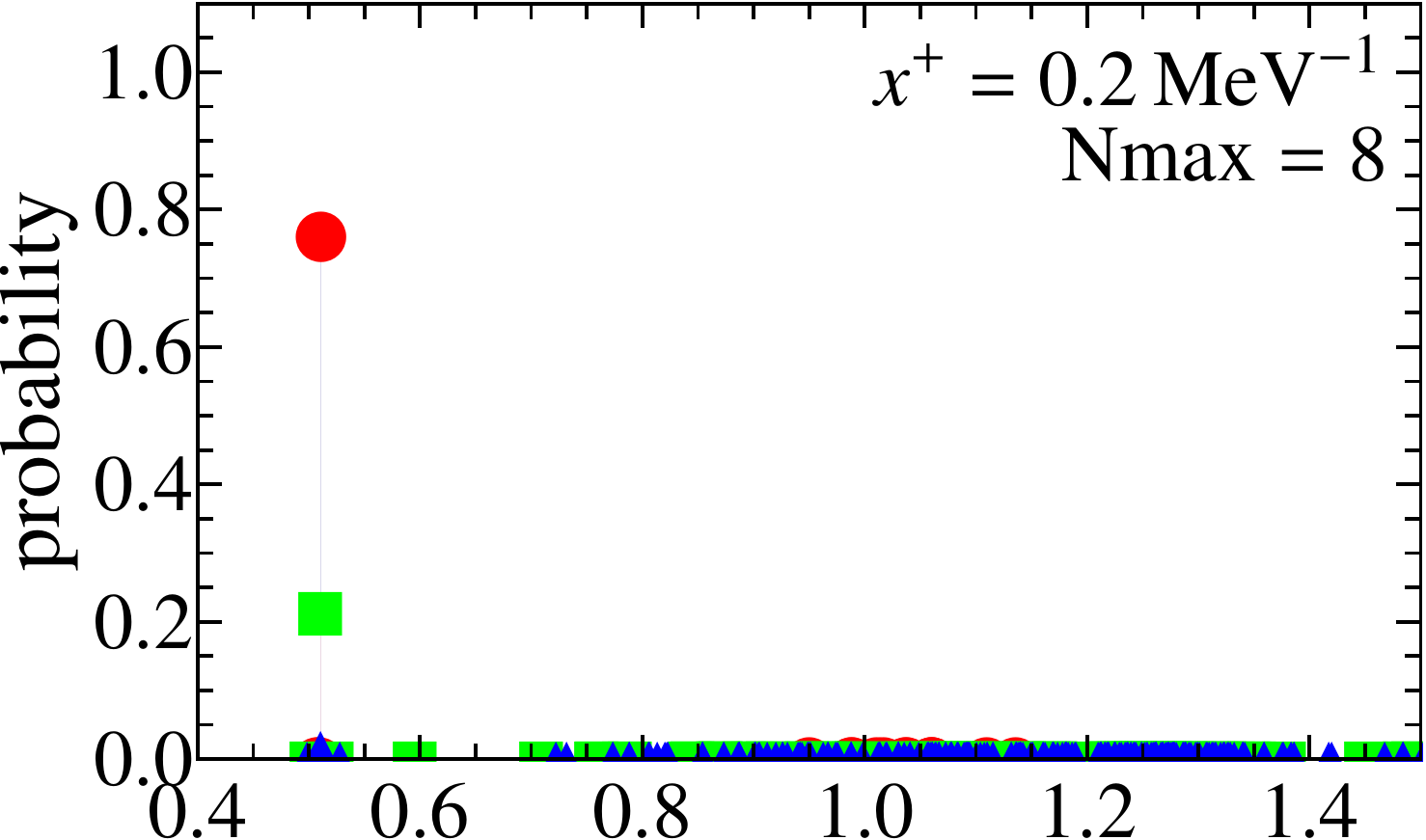}
\includegraphics[width=0.46\textwidth]{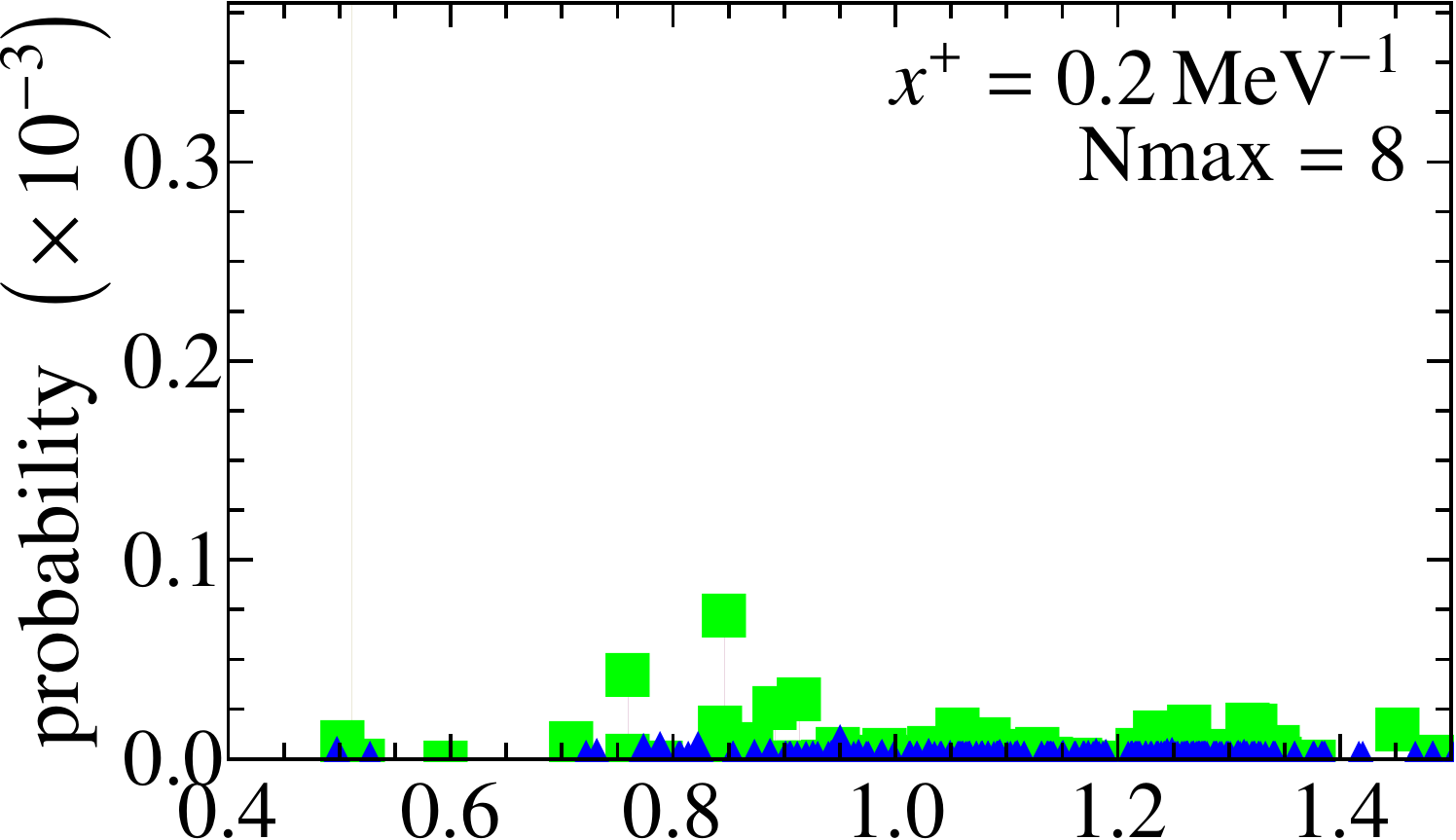}
\includegraphics[width=0.45\textwidth]{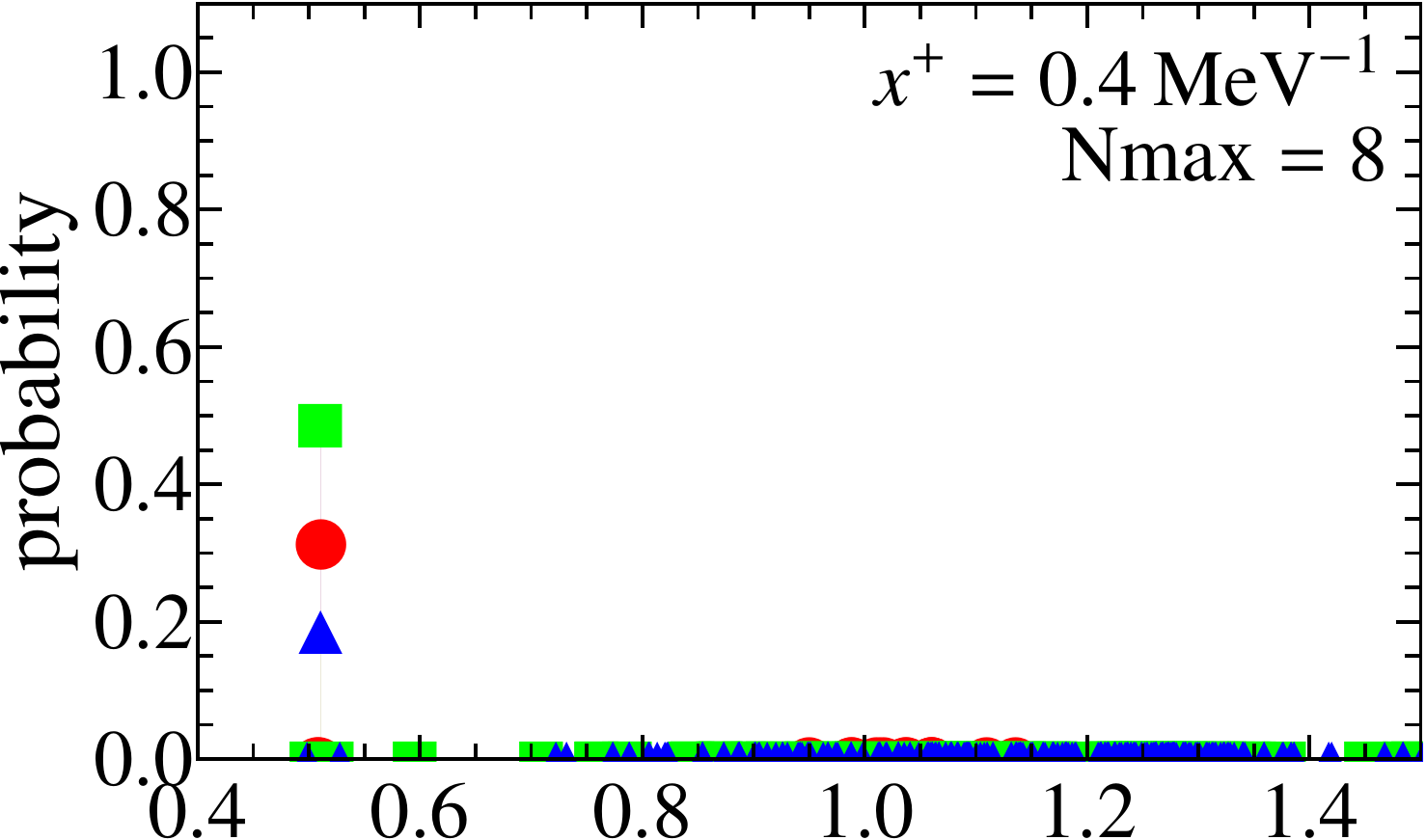}
\includegraphics[width=0.46\textwidth]{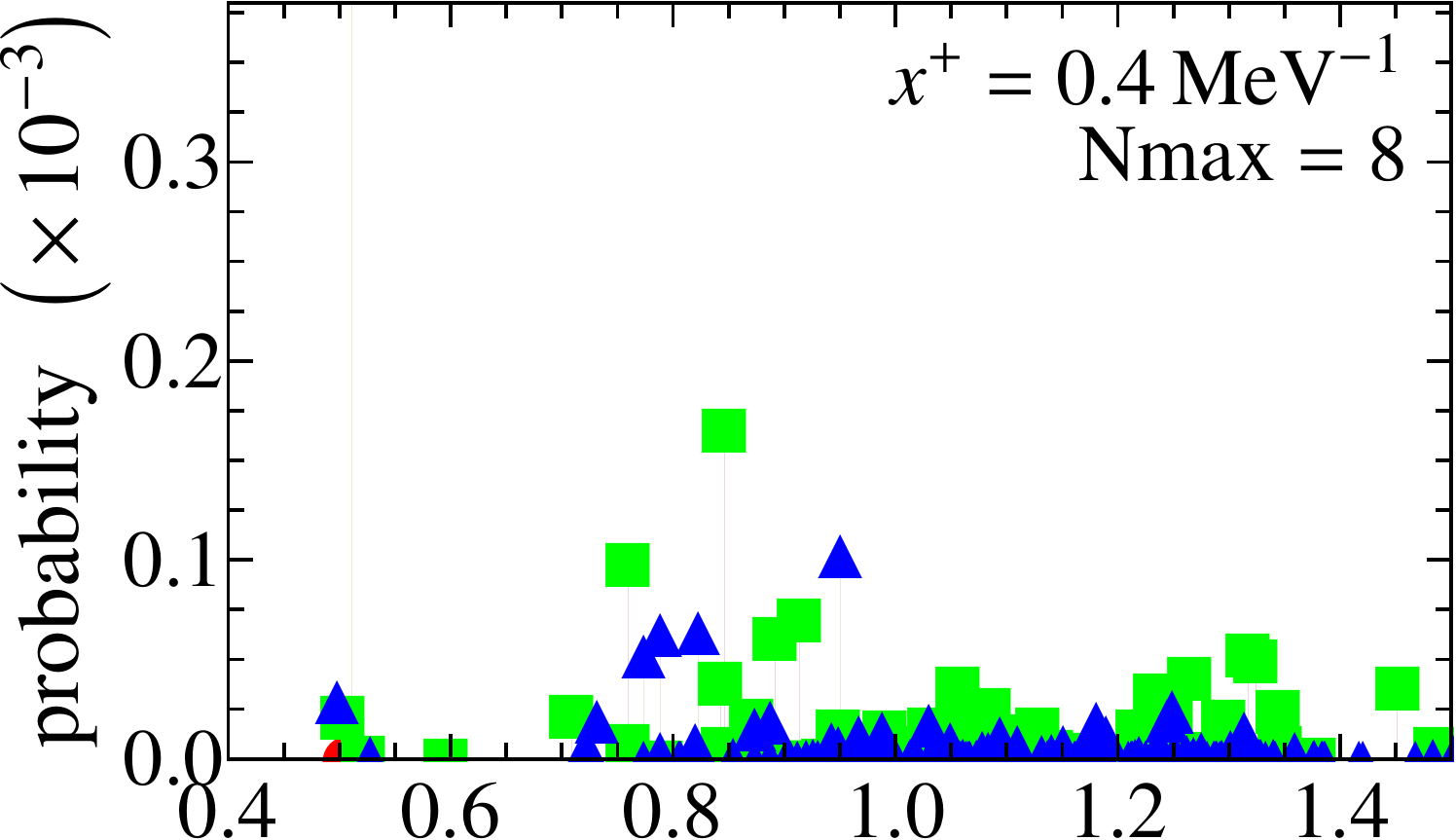}
\includegraphics[width=0.45\textwidth]{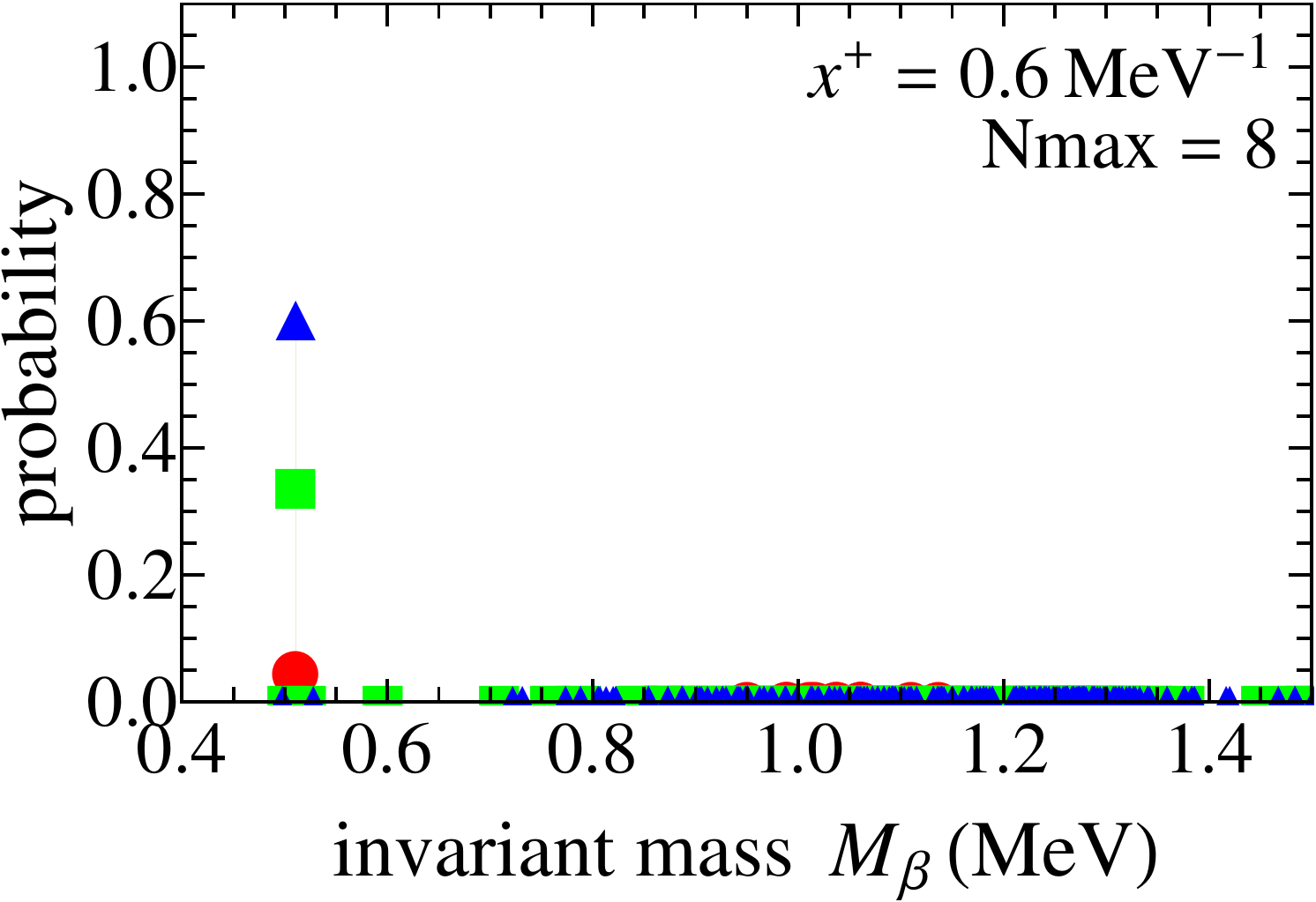}
\includegraphics[width=0.46\textwidth]{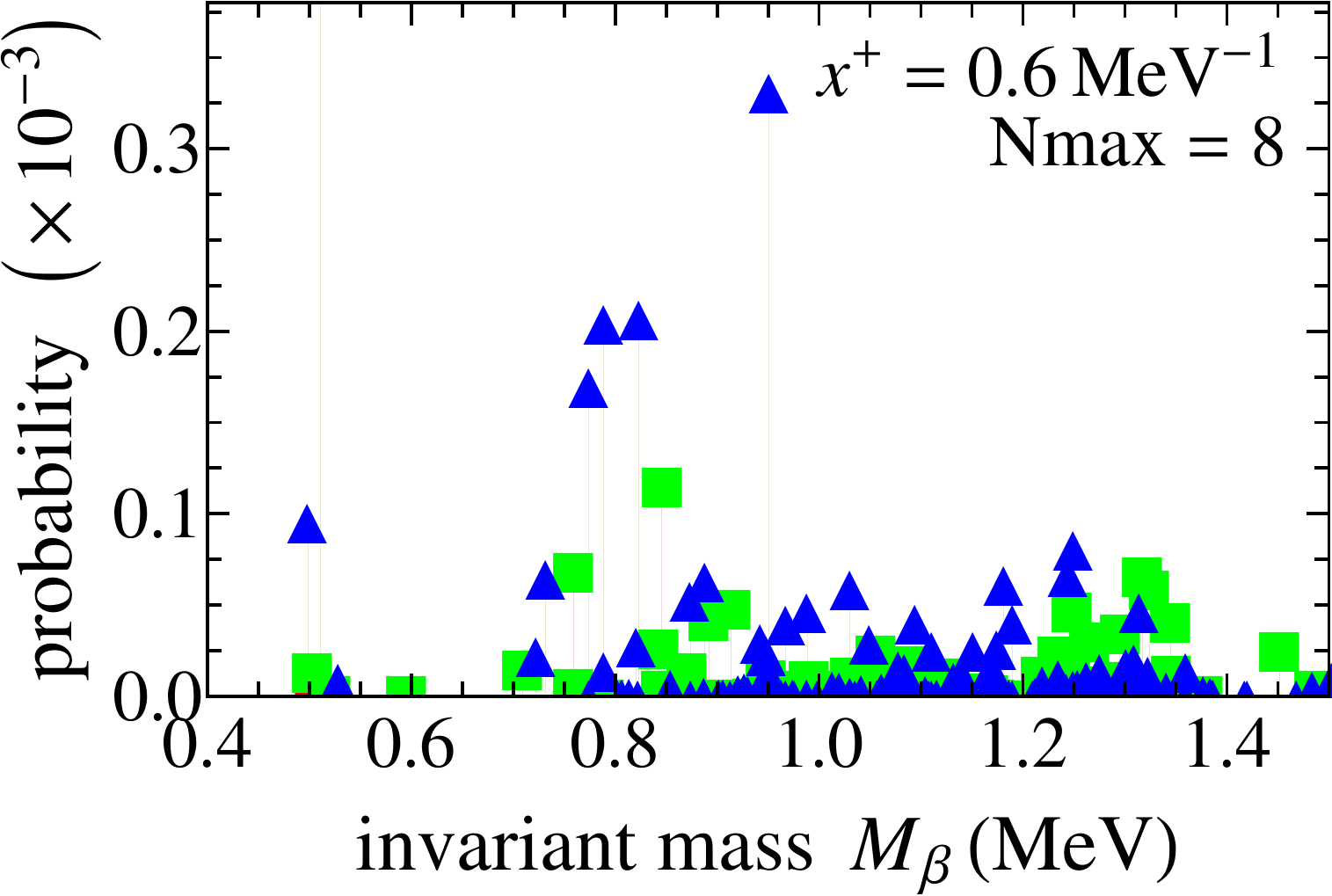}

\caption{\label{fig:state_evol_np}  (Color online) Time evolution of the single electron system in the laser field. From top to bottom, the panels in each row successively correspond to lightfront-time $x^+$=0, 0.2, 0.4, 0.6MeV$^{-1}$ (the laser field is switched on at $x^+$=0). Each dot on these plots stands for a tBLFQ basis state. Y-axis is the probability for the tBLFQ basis state $|c_\beta(x^+)|^2$ and x-axis is its corresponding invariant mass $M_{\beta}$. The panels on the left (with y-axis up to 1.1) illustrate the evolution of the single electron (ground) states in $K$=1.5, 3.5, 5.5 segments respectively and the panels on the right with y-axis ``zoomed-in'' show the evolution of various electron-photon (excited) states. The electromagnetic coupling constant $\alpha{=}e^2/(4\pi)$ is 1/137.}
\end{figure*}

With the laser matrix elements checked, we now turn to nCs in a larger basis. This basis consists of three segments with $K{=}\{K_i, K_i{+}k_\text{las}, K_i{+}2k_\text{las}\}$. In each segment we retain {\it both} the single electron (ground) and electron-photon (excited) state(s). The initial state for the nCs process is a single (ground state) electron in the $K{=}K_i$ segment. This basis allows for the ground state to be excited twice by the background (from the segment with $K$=$K_i$ through to segment with $K_i$+2$k_\text{las}$). In this calculation, we take $K_i{=}1.5$ and $N_\text{max}{=}8$, with $a_0=10$ and $k_\text{las}{=}2$. We present the evolution of the electron system in Fig.~\ref{fig:state_evol_np}, at increasing (top to bottom) lightfront time.
The initial system is shown in the top panel of Fig.~\ref{fig:state_evol_np}; the only populated basis state is the single electron (ground) state in the $K{=}1.5$ segment. As time evolves, the background causes transitions from the ground state to states in the $K{=}3.5$ segment. Both the single electron state and electron-photon states are populated; the former represent the acceleration of the electron by the background, while the later represent the process of radiation. At times $x^+{=}0.2$~MeV$^{-1}$, the single electron state\footnote{Because we neglect counter-terms, the single electron ground states in the $K{=}1.5$, $3.5$ and $5.5$ segments receive increasing (negative) mass corrections from loop effects. $K$ works as an ultraviolet and infrared regulator in the longitudinal direction (see discussion in Sect.~\ref{3b}) and as a result, the calculated value for the invariant mass of the $K{=}3.5$ and $K{=}5.5$ single electron states is slightly lower than that for $K{=}1.5$. In order to prevent the invariant mass of the whole system being affected by this artifact, we manually set the invariant mass for each $K$-segment single electron state to the physical mass $m_e$.} in $K{=}3.5$ becomes populated while the probability for finding the initial state begins to drop. In the right hand panel, the populated electron-photon states begin forming a peak structure. The location of the peak is around the invariant mass of 0.8~MeV, roughly consistent with the expected value of $M_\text{pk1}{=}\sqrt{P^-(K_i+k_\text{las})}{=}0.78$~MeV, cf. the discussion in Section~\ref{ssec:cali}. 

Once the basis states in $K{=}3.5$ become populated, ``second" transitions to the $K{=}5.5$ segment become possible. This can be seen in the third row of Fig.~\ref{fig:state_evol_np}, at $x^+{=}0.4$~MeV$^{-1}$. In the left hand panel, one sees that the probability of the electron to remain in its ground state ($K{=}1.5$) is further decreased, the probability of it being accelerated (to $K{=}3.5$) is increased, and that the $K{=}5.5$ single electron state becomes populated. In the right hand panel, the electron-photon states in the $K{=}5.5$ segment also become populated as a result of the second transitions. A second peak arises here at the invariant mass of around $M_\text{pk2}{=}\sqrt{P^-(K_i+2k_\text{las})}{\sim}1.0$~MeV (distinct from at that ${\sim}0.8$~MeV, above, formed by the $K{=}3.5$ electron-photon states from the first transitions). The peak in the $K{=}5.5$ segment is at a larger invariant mass than that in the $K{=}3.5$ segment simply because the basis states in the $K{=}5.5$ segment follow from the initial state being excited {\it twice} by the background field, and thus receive more energy than states in the $K{=}3.5$ segment.

As time evolves further, the probability of finding a $K{=}5.5$ single electron exceeds that of finding a $K{=}3.5$ segment electron, see the bottom left panels in Fig.~\ref{fig:state_evol_np}. At this time, $x^+{=}0.6$~MeV$^{-1}$, the system is most likely be found in the $K{=}5.5$ single electron state, with probability $\sim$0.6. The probability for finding the $K{=}3.5$ single electron state is around~$0.35$ and the initial $K{=}1.5$ electron state is almost completely depleted. In the right hand panel, we see that the probability for finding $K$=5.5 electron-photon states increases with time. One also notices that at later times, the probability for $K$=3.5 electron-photon states also begin to drop. This is because (like the $K$=3.5 single electron) the $K{=}3.5$ electron-photon states are coupled to the $K{=}5.5$ single electron; as the probability of the $K$=5.5 single electron state increases, it ``absorbs" both the single electron and the electron-photon states in the $K$=3.5 segment. At $x^+$=0.6MeV$^{-1}$ we terminate the evolution process, as the system is already dominated by the single electron state in the maximum $K$-segment. Further evolution without artifacts would require bases with segments of $K{=}\{7.5, 9.5\ldots\}$.

This calculation, although performed in a basis of limited size, illustrates the basic elements of the tBLFQ framework. The acceleration of the single electron state and the radiation of a photon are treated coherently within the same Hilbert space.

\begin{figure}[!t]
\centering
\includegraphics[width=0.47\textwidth]{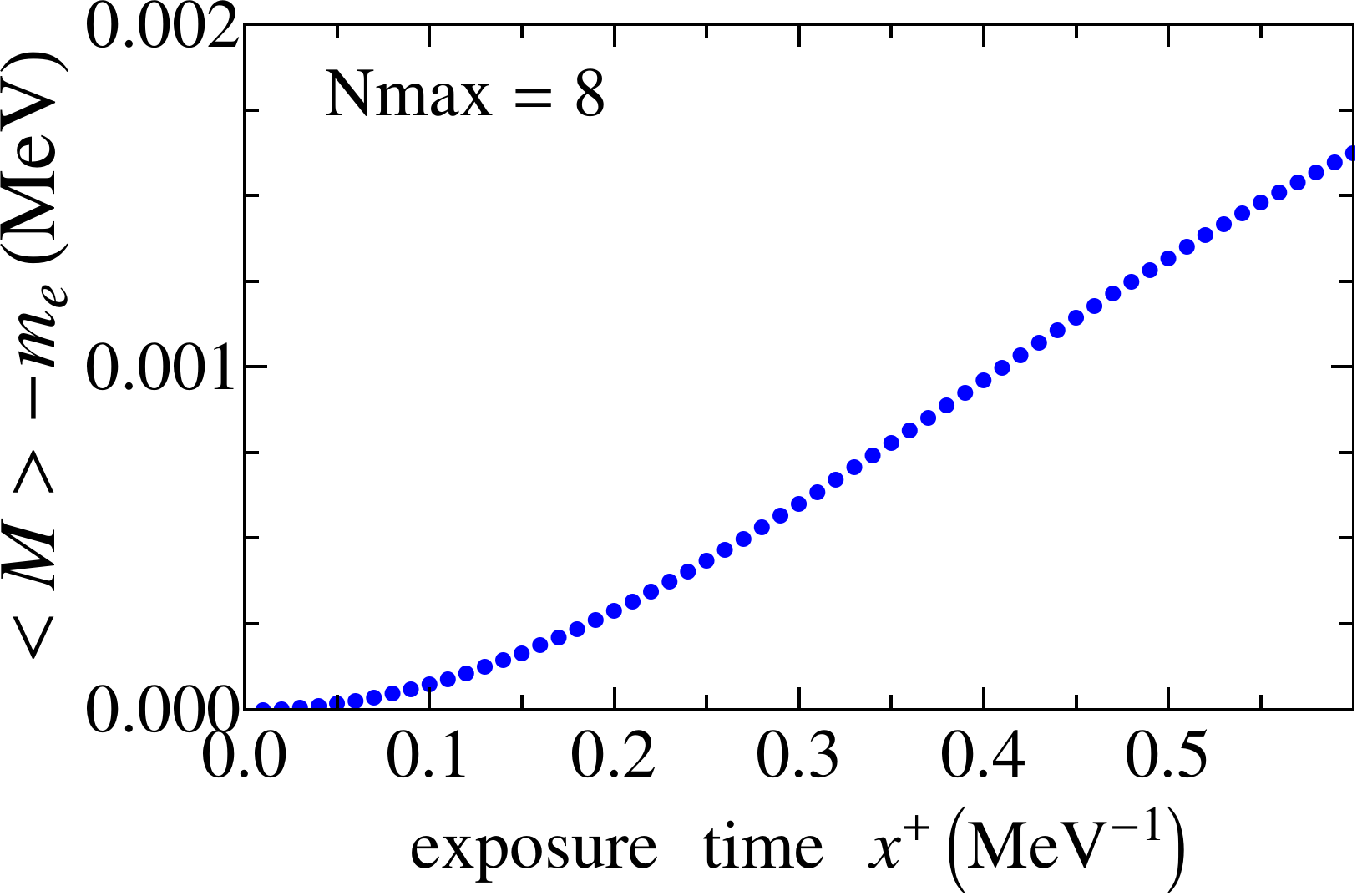}
\caption{(Color online) Time evolution of the average invariant mass of the electron system calculated in tBLFQ basis space with $N_\text{max}$=8. Y-axis is the difference between the average invariant mass $\langle M\rangle$ of the system at $x^+$ and that of a single electron $m_e$. X-axis is the (lightfront) exposure time $x^+$. The electromagnetic coupling constant $\alpha{=}e^2/(4\pi)$ is 1/137.}
\label{fig:state_evol_invmass}
\end{figure}

Since the states $\ket{\beta}$ encode all the information of the system, they can be employed to construct other observables. As an example, in Fig.~\ref{fig:state_evol_invmass} we present the evolution of the average invariant mass $\langle M\rangle$ of the system as a function of time. The increase of the invariant mass with time reflects the fact that energy is pumped into the electron-photon system by the laser field.  This invariant mass can be accessed experimentally by measuring the momenta of both the final electron, $p^\mu_e$, and photon, $p^\mu_\gamma$ in an nCs experiment. The invariant mass can be compared with the expectation value of $(p^\mu_e+p^\mu_\gamma)^2$ measured over many repetitions of the nCs experiment.
Work in deriving other observables, such as the cross sections for specific electron-photon final states, is in progress.

In this section we have demonstrated a) the general procedure for treating processes  nonperturbatively in tBLFQ, and b) the accessibility of the full configuration (wavefunction) of the system at finite time.
 
\section{Conclusions and Outlook}
\label{sec:concl}
In this paper we constructed a nonperturbative framework for time-dependent problems in quantum field theory, referred to as
time-dependent BLFQ (tBLFQ). This framework is based on the previously developed Basis Light-front Quantization (BLFQ) and adopts
the light-front Hamiltonian formalism. Given the Hamiltonian and the initial configuration of a quantum field system as input, the
system's subsequent evolution is evaluated by solving the Schr\"odinger equation of light-front dynamics. The eigenstates of the
time-independent part of the Hamiltonian, found by the BLFQ approach, provide the basis for the time-evolution process. Basis
truncation and time-step discretization are the only approximations in this fully nonperturbative approach. (Note that the choice of background field is an input parameter; although a simple background is adopted in this work, the tBLFQ framework is in principle capable of dealing with realistic background fields with generic spatial and temporal dependence.) One feature of the tBLFQ framework is that the complete wavefunction of the quantum field system is accessible at any intermediate time during the evolution, which provides convenience for detailed studies of time-dependent processes.

As an initial application we have applied this framework to an external field problem. We have studied the process in which an electron absorbs energy-momentum from an intense background laser field, and emits a single photon. In contrast to current numerical approaches to strong laser physics, tBLFQ is fully quantum mechanical and allows us to see both the acceleration of the electron by the background and the creation of a photon, in real-time. Note that tBLFQ is also applicable to problems without external fields but in which nontrivial time-dependence arises from using an initial state which is a non-stationary superposition of mass eigenstates.

Future developments will be made in two directions. First, further improvement of tBLFQ itself. The initial step is to implement renormalization so that the BLFQ representation of the physical eigenspectrum of QED can be improved (and then used in tBLFQ calculations). Currently we are working on implementing a sector-dependent renormalization scheme within the BLFQ framework. The inclusion of higher Fock sectors in our calculation is also important, as it will not only result in more realistic representations of quantum states but will also allow for the description of a larger variety of processes, \eg, multi-photon emissions. 

The second direction to be pursued is the extension of tBLFQ's range of applications. In the field of intense laser physics, the inclusion of transverse ($x^\perp$), longitudinal ($x^-$) and time ($x^+$) dependent structures to the background field will be used to more realistically model the focussed beams of next-generation laser facilities~\cite{IVAN}.  In addition to intense laser physics, we will also apply tBLFQ to relativistic heavy-ion physics, specifically the study of particle production in the strong (color)-electromagnetic fields of two colliding nuclei. Ultimately, the goal is to use tBLFQ to address strong scattering problems with hadrons in the initial and/or final states. As supercomputing technology continues to evolve, we envision that tBLFQ will become a powerful tool for exploring QCD dynamics.

\acknowledgments

We acknowledge valuable discussions with K. Tuchin, H. Honkanen, S. J. Brodsky, P. Hoyer, P. Wiecki and Y. Li. This work was supported in part by the Department of Energy under Grant Nos. DE-FG02-87ER40371 and DESC0008485 (SciDAC-3/NUCLEI) and by the National Science Foundation under Grant No. PHY-0904782. A.~I.\ is supported by the Swedish Research Council, contract 2011-4221. Diagrams created using JaxoDraw \cite{Binosi:2003yf,Binosi:2008ig}.

\appendix

\section{The light-front QED Hamiltonian}
\label{hami_full}
%
In this section we follow the derivation of the Hamiltonian in \cite{Brodsky:1997de}, but with an additional background field. The Lagrangian is
\be
	\mathcal{L} = -\frac{1}{4}F_{\mu\nu}F^{\mu\nu} + \bar{\Psi}(i\gamma^\mu D_\mu-m_e)\Psi \;,
\ee
in which $D_\mu \equiv \partial_\mu +ie C_\mu$ and $C_\mu = \mathcal{A}_\mu + A_\mu$ is the sum of the background and quantum gauge fields respectively. Note that $F_{\mu\nu}$ is calculated from $A_\mu$ alone, \ie\ there is no kinetic term for the background. The equations of motion for the fields are
\be\label{EOM-GAUGE}
	\partial_\mu F^{\mu\nu} = e\bar{\Psi}\gamma^\nu\Psi \equiv e j^\nu\;,
\ee
which defines the current $j^\nu$, and
\be\label{EOM-SPINOR}
	\big[i\gamma^\mu D_\mu-m_e\big]\Psi =0 \;.
\ee
The background field appears in the equations of motion for the fermion, but not for the gauge field. We now analyze these equations in light-front coordinates ($x^\pm=x^0\pm x^3$, and $x^\pm = 2x_\mp$).  We work in light-front gauge, so that $A^\LCp=\mathcal{A}^\LCp=0$.  The $\nu=+$ component of (\ref{EOM-GAUGE}) does not contain time derivatives, and can be written
\be\label{CONSTRAINT-GAUGE}
	\frac{1}{2}  A^\LCm = \frac{\partial^\LCperp A^\LCperp}{\partial^\LCp} - e \frac{j^\LCp}{(\partial^\LCp)^2} \;.
\ee
This is a constraint equation which relates the (non-dynamical) field $A^\LCm$ to the transverse components $A^\LCperp$ and the fermion current. Similarly, if we multiply (\ref{EOM-SPINOR}) by $\gamma^\LCp$ on the left, we find a constraint equation for the fermion field. Defining first the orthogonal field components
\be
	\Psi_\LCm \equiv \tfrac{1}{4}\gamma^\LCp\gamma^\LCm\Psi \;,\qquad \Psi_\LCp \equiv \tfrac{1}{4}\gamma^\LCm\gamma^\LCp\Psi \;,
\ee
the constraint equation may be written
\be\label{CONSTRAINT-SPINOR}
	\Psi_\LCm = \frac{1}{2i\partial^\LCp}\big[m_e-i\gamma^\LCperp D^\LCperp\big]\gamma^\LCp \Psi_\LCp \;.
\ee
Hence, the field $\Psi_\LCm$ is non-dynamical and can be expressed in terms of the dynamical field $\Psi_\LCp$.
We now turn to the construction of the Hamiltonian. The conjugate momentaü are
\be
	\frac{\partial \mathcal{L}}{\partial \partial_\LCp\Psi} = i\bar\Psi \gamma^\LCp \;, \qquad \frac{\partial\mathcal{L}}{\partial\partial_\LCp A_\mu} = F^{\mu+}
\ee
and the Hamiltonian $P^\LCm = 2P_\LCp$ is then
\be\begin{split}
	P^\LCm &= \int\!\ud^2x^\LCperp\ud x^\LCm\  F^{\mu+}\partial_+ A_\mu + i\bar\Psi \gamma^\LCp \partial_\LCp\Psi - \mathcal{L} \\
	&= \int\!\ud^2x^\LCperp\ud x^\LCm\  F^{\mu+}\partial_+ A_\mu + \frac{1}{4}F^{\mu\nu}F_{\mu\nu} + i\bar\Psi \gamma^\LCp \partial_\LCp\Psi \;,
\end{split}
\ee
in which the first line is the standard Legendre transformation, and in the second line we have used the equations of motion. It is convenient to add a total derivative to the Hamiltonian \cite{Brodsky:1997de}, the term $-\partial_\mu(F^{\mu+}A_\LCp)$, and again use the equations of motion to write
\be\begin{split}\label{PROTO}
	P^\LCm = \int\!\ud^2 x^\LCperp\ud x^\LCm\  &\frac{1}{4}F^{\mu\nu}F_{\mu\nu} - F^{\mu+}F_{\mu+} \\
	&+ i\bar\Psi \gamma^\LCp D_\LCp\Psi  +e \bar{\Psi}\gamma^\LCp \mathcal{A}_\LCp \Psi\;.
\end{split}
\ee
In order to complete the transition to the Hamiltonian picture we need to eliminate the light-front time derivatives of the fields in favour of the fields themselves, and their momenta. The gauge field terms are simplest. Let $i,j$ be transverse indices and define
\be
	\{\tilde{A}^\LCp,\tilde{A}^-,\tilde{A}^j\}:= \{0,2\frac{\partial^jA^j}{\partial^\LCp},A^j\} \;.
\ee
The first line of (\ref{PROTO}) then becomes
\be\begin{split}
	&\frac{1}{4}F^{ij}F_{ij} - \frac{1}{2}F^{+-}F_{+-} \\
	&= \frac{1}{2} {\tilde A}^j (i\partial^\LCperp)^2 {\tilde A}^j + \frac{e^2}{2} j^\LCp \frac{1}{(i\partial^\LCp)^2}j^\LCp + e j^\LCp \tilde{A}_\LCp \;,
\end{split}	
\ee
using the constraint (\ref{CONSTRAINT-GAUGE}). The field $\tilde{A}_\mu$ is that which survives the limit $e\to 0$, and is therefore referred to as a ``free field''. Turning now to the spinor terms in (\ref{PROTO}), we have
\be\label{SPIN-BIDRAG}
	 i\bar\Psi \gamma^\LCp D_\LCp\Psi = 2i \Psi^\dagger_\LCp D_\LCp \Psi_\LCp \;,
\ee
and the spinor equations of motion (\ref{EOM-SPINOR}) then give
\be\begin{split} \label{yawn}
	2i  D_\LCp \Psi_\LCp &=  \frac{1}{2}[m_e-i\gamma^\LCperp D^\LCperp]\gamma^\LCm \Psi_\LCm \\
	&= \frac{1}{2}[m_e-i\gamma^\LCperp D^\LCperp] \frac{\gamma^\LCm}{2i\partial^\LCp}  [m_e-i\gamma^\LCperp D^\LCperp] \gamma^\LCp \Psi_\LCp\\
	&=  [m_e-i\gamma^\LCperp D^\LCperp] \frac{1}{i\partial^\LCp}  [m_e+i\gamma^\LCperp D^\LCperp] \Psi_\LCp \,.
\end{split}
\ee
The first line follows from $\gamma^\LCm\Psi_\LCp\equiv 0$, we used (\ref{CONSTRAINT-SPINOR}) in the second line and in the third line we commuted $\gamma^\LCm$ to the right. In analogy to $\tilde{A}$, we introduce $\tilde{\Psi}$, defined by
\be\label{TILDE-DEF}
	\tilde{\Psi}_\LCp = {\Psi}_\LCp \;,  \qquad  \tilde{\Psi}_\LCm = \frac{1}{2i\partial^\LCp} \big[ m_e - i\gamma^\LCperp \partial^\LCperp\big]\gamma^\LCp \tilde{\Psi}_\LCp \;.
\ee
Again, this is the field which survives the $e\to 0$ limit. Our final task is to insert (\ref{TILDE-DEF}) into (\ref{yawn}) and rewrite this in terms of only the ``tilde'' variables. First, the $C$-free terms of of (\ref{SPIN-BIDRAG}) are:
\be\label{A-0}
	\Psi^\dagger_\LCp [m_e-i\gamma^\LCperp \partial^\LCperp] \frac{1}{i\partial^\LCp}  [m_e+i\gamma^\LCperp \partial^\LCperp] \Psi_\LCp = \frac{1}{2}\bar{\tilde{\Psi}} \gamma^\LCp \frac{m^2_e+(i\partial^\LCperp)^2}{i\partial^\LCp}\tilde\Psi \;.
\ee
Next, we have terms in (\ref{SPIN-BIDRAG}) which are linear in $C$:
\be\begin{split}\label{A-LIN}
	\Psi^\dagger_\LCp [&e\gamma^\LCperp C^\LCperp] \frac{1}{i\partial^\LCp}  [m_e+i\gamma^\LCperp \partial^\LCperp] \Psi_\LCp \\
	&+ \Psi^\dagger_\LCp  [m_e-i\gamma^\LCperp \partial^\LCperp] \frac{1}{i\partial^\LCp} [-e\gamma^\LCperp C^\LCperp] \Psi_\LCp \\
	&= \frac{1}{2}\tilde{\Psi}^\dagger_\LCp [e\gamma^\LCperp C^\LCperp] \gamma^\LCm \tilde{\Psi}_\LCm + \frac{1}{2}\tilde{\Psi}^\dagger_\LCm \gamma^\LCp [-e\gamma^\LCperp C^\LCperp] \tilde{\Psi}_\LCp \\
	&=  e\tilde{j}^\LCperp C_\LCperp \;, \\ \vspace{1pt}
\end{split}
\ee
using (\ref{TILDE-DEF}) in the second line. Note the tilde on $j$ in the third line.  Finally, we have the terms quadratic in $C$, which are
\be\label{A-QUAD}
	-e^2 \tilde{\Psi}_\LCp[e\gamma^\LCperp C^\LCperp] \frac{1}{i\partial^\LCp} [e\gamma^\LCperp C^\LCperp] \tilde\Psi_\LCp \;.
\ee
Now, we sum (\ref{yawn}) (\ref{A-0}), (\ref{A-LIN}) and (\ref{A-QUAD}) to obtain the full Hamiltonian: we {\it drop} the ``tilde'' on all variables from now on, so that one must remember that
\be
	A^\LCm \equiv 2\frac{\partial^\LCperp A^\LCperp}{\partial^\LCp} \;, \qquad \Psi_\LCm \equiv \frac{1}{2i\partial^\LCp} \big[ m_e - i\gamma^\LCperp \partial^\LCperp\big]\gamma^\LCp \Psi_\LCp \;.
\ee
Since we are interested in the new interactions introduced by the background field, we will separate these out explicitly, expanding $C\to \mathcal{A} + A$. (Recall, $A$ has a tilde now.) Finally, the full Hamiltonian is
\begin{widetext}
\be\begin{split}\label{FULL}
	P^- = \int\!\ud^2x^\LCperp\ud x^\LCm \  &\frac{1}{2}\bar{{\Psi}} \gamma^\LCp \frac{m_e^2+(i\partial^\LCperp)^2}{i\partial^\LCp}\Psi + \frac{1}{2} { A}^j (i\partial^\LCperp)^2 { A}^j  + e{j}^\mu {A}_\mu  + \frac{e^2}{2} { j}^\LCp \frac{1}{(i\partial^\LCp)^2}{ j}^\LCp + \frac{e^2}{2}\, \bar{\Psi} \gamma^\mu { A}_\mu \frac{\gamma^\LCp}{i\partial^\LCp} \gamma^\nu {A}_\nu \Psi  \\
	&+ ej^\mu \mathcal{A}_\mu + \frac{e^2}{2}\, \bar{\Psi} \gamma^\mu \mathcal{A}_\mu \frac{\gamma^\LCp}{i\partial^\LCp} \gamma^\nu \mathcal{A}_\nu \Psi +\frac{e^2}{2}\, \bar{\Psi} \gamma^\mu { A}_\mu \frac{\gamma^\LCp}{i\partial^\LCp} \gamma^\nu \mathcal{A}_\nu \Psi + \frac{e^2}{2}\, \bar{\Psi} \gamma^\mu \mathcal{A}_\mu \frac{\gamma^\LCp}{i\partial^\LCp} \gamma^\nu {A}_\nu \Psi \;.
	\end{split}	
\ee
\end{widetext}
The first line is the QED light-front Hamiltonian, $P^-_\text{QED}$. The second line contains the new terms generated by the
background field. We label the terms in $P^-_\text{QED}$ as $T_f$, $T_\gamma$, $W_1$\ldots $W_3$ respectively. $T_f$ and
$T_\gamma$ are the kinetic energy terms for the fermion and gauge field respectively. $W_1$ is called the {\it vertex
  interaction}, which is responsible for photon emission and electron-positron pair-production processes. $W_2$ is the {\it
  instantaneous-photon interaction} and $W_3$ is the {\it instantaneous-fermion interaction}. The instantaneous-photon interaction
is the light-front analogue of the Coulomb energy, and its origin is Gauss's law~(\ref{CONSTRAINT-GAUGE})
\cite{Brodsky:1997de}. The instantaneous-fermion interaction is (explicitly) present exclusively in light-front
dynamics. In perturbation theory, methods to properly treat the IR divergences associated with these instantaneous
  interactions, have been developed and applied in Refs \cite{Zhang:1993is, Valdes:2004ew}.

The additional terms introduced by the background field, in the second line of (\ref{FULL}), are the three-point background vertex interaction, the instantaneous  2-background, 2-fermion vertex, and two, instantaneous, 1-background, 1-photon, 2-fermion vertices. For the field (\ref{laser-profile}), the Hamiltonian (\ref{FULL}) contains only a single term beyond the ordinary QED Hamiltonian, this term being
\be
	j^\LCp \mathcal{A}_\LCp = \bar{\Psi}\gamma^\LCp \Psi \mathcal{A}_\LCp = 2\Psi^\dagger_\LCp \Psi_\LCp \mathcal{A}_\LCp = \Psi^\dagger_\LCp \Psi_\LCp \mathcal{A}^\LCm \;. 
\ee
The spatial integral of this term is denoted as $V$ in Sec. IV and V, and in Appendix E.  As the main goal in this work is to introduce the general framework of BLFQ (rather than to present new results through precise numerical calculations, for which see future articles) we work for convenience with a truncated QED Hamiltonian, dropping the instantaneous interaction terms $W_2$ and $W_3$, proportional to $e^2$. The remaining Hamiltonian is sufficient for calculating eigenstates and eigenvalues to first order in $\alpha$.

\subsection{Symmetries of $P^-_\text{QED}$}
The BLFQ basis explicitly carries three of the symmetries of the QED light-front Hamiltonian $P^-_\text{QED}$. A fourth symmetry, boost invariance in the transverse direction, is not encoded directly in the basis.  However, as discussed in the text, due to the choice of HO basis and the $N_\text{max}$ truncation method, this symmetry is recovered by a factorization of the resulting amplitudes into a component for center-of-mass motion times the components for internal motion~\cite{Zhao:2013xx}. The three symmetries encoded directly in the basis and their operators, which commute with $P_\text{QED}^-$, are listed below.

The longitudinal momentum operator is,
\begin{align}
    \label{longitudinal_momentum_operator}
    P^+=\frac{1}{2}\int\! \ud x^\LCm \ud^2 x^\LCperp\ 2\Psi^\dagger_+i\partial^+\Psi_++\partial^+A^j\partial^+A^j\ ,
\end{align} 
with $j\in\{1,2\}$. This commutes with $P^-_\text{QED}$, and so overall longitudinal momentum is conserved.

The longitudinal projection of angular momentum is also conserved. The corresponding operator is written $J^3$. This can be decomposed into the following four parts,
\begin{align}
    \label{longitudinal_projection_angular_momentum_decomposition}
    J^3=J^3_{f;o}+J^3_{f;i}+J^3_{\gamma;o}+J^3_{\gamma;i}
\end{align}
in which the subscript ``$o$'' refers to the longitudinal projection of orbital angular momentum, while subscript ``$i$'' refers to the longitudinal projection of the spin angular momentum, \ie\ helicity. The subscripts $f$ and $\gamma$ refer to the fermion and photon, as above. In terms of the fields, these four operators are
\begin{align}
    &J^3_{f;o} = \int\!\ud x^\LCm\ud^2x^\LCperp\Psi^\dagger_+i(x^1\partial^2-x^2\partial^1)\Psi_+\ ,\nonumber\\
    &J^3_{f;i} = \int\!\ud x^\LCm\ud^2x^\LCperp\ \Psi^\dagger_+\Sigma^3\Psi_+\ ,\nonumber\\
    &J^3_{\gamma;o}=\frac{1}{2}\int\!\ud x^\LCm\ud^2x^\LCperp\ x^1[\partial^+A^1\partial^2A^1+\partial^+A^2\partial^2A^2 \nonumber\\& \qquad\qquad{}-x^2[\partial^+A^1\partial^1A^1+\partial^+A^2\partial^1A^2]\ ,\nonumber\\
    &J^3_{\gamma;i}=\frac{1}{2}\int\!\ud x^\LCm\ud^2x^\LCperp A^1\partial^+A^2-A^2\partial^+A^1\ .
\end{align}
where $\Sigma^3\equiv\begin{pmatrix}\sigma^3&0\\0&\sigma^3\end{pmatrix}$. 

Finally, net fermion number is also conserved. The corresponding operator is
\begin{align}
    N_f=\int dx^-d^2x^\perp \Psi^\dagger_+ \Psi_+\ .
\end{align}
%

%
\section{BLFQ harmonic oscillator basis}
\label{Transverse}
%
Our BLFQ basis elements differ from the usual basis of momentum states only in the transverse degrees of freedom. In this appendix, we describe the transverse structure of our basis elements.

For the transverse part, the basis elements are eigenstates of the following two-dimensional harmonic oscillator (2D-HO) Hamiltonian 
\begin{align}
    \label{HO_hami}
    H^{2d}_{HO}=\frac{p^2_\perp}{2M}+\frac{1}{2}M\Omega^2 x^2_\perp,
\end{align}
in which $M$ and $\Omega$ are the mass and frequency of the oscillator. The choice of these free parameters will be discussed shortly. The characteristic scale of the 2D-HO is $b=\sqrt{M\Omega}$ will be called the 2D-HO parameter. The eigenstates of (\ref{HO_hami}) are labelled by two quantum numbers, $n$, the principle quantum number characterizing the quanta of the radial excitation, and $m$, the angular quantum number characterizing the angular momentum. These eigenstates, which we write $\ket{nm}$ have eigenenergy $E_{n,m}=(2n+|m|+1)\Omega$. In coordinate space, the corresponding wavefunctions can be factorized into a conventional angular part $\chi_m(\phi)$,
\begin{align}
    \chi_m(\phi)=\frac{1}{\sqrt{2\pi}}e^{im\phi}\ ;
\end{align}
and a radial part $f_{nm}(\rho)$, as follows
\begin{align}
    \label{transverse_HO_wavefunction}
    \Phi^b_{nm}(\rho,\phi)=\braket{x^\perp}{nm}=(-1)^{n}i^{|m|}f_{nm}(\rho)\chi_m(\phi) \;,
\end{align}
where ($\rho$, $\phi$) are polar coordinates in the transverse plane, $x^1=\rho \cos\phi$ and $x^2=\rho \sin\phi$. Explicitly, the radial part $f_{nm}(\rho)$ is given in terms of generalized Laguerre polynomials, $L^{|m|}_n(b^2\rho^2)$, by
\begin{align}
    f^b_{nm}(\rho)=& b\sqrt{2}\sqrt{\frac{n!}{(n+|m|)!}}\ e^{-b^2\rho^2/2} (b\rho)^{|m|}L^{|m|}_n(b^2\rho^2)\; ,
\end{align}
The 2D-HO wavefunctions $\Phi^b_{nm}(\rho,\phi)$ satisfies the following orthonormalization condition,
\begin{align}
    \braket{nm}{n'm'}&=\int^\infty_0\int^{2\pi}_0\! \ud\rho\rho\, \ud \phi\ \Phi^{b*}_{nm}(\rho,\phi)\Phi^b_{n'm'}(\rho,\phi)\nonumber\\
    &=\delta^{n'}_{n}\delta^{m'}_{m} \;.
\end{align}
In general, the 2D-HO wavefunction $\Phi^b_{nm}(\rho,\phi)$ is a highly oscillatory function with respect to both $\rho$ and $\phi$. In the radial direction, though, the oscillations terminate with a steep fall-off to zero at around $b\rho \sim 2\sqrt{2n+|m|}$.

One property of these wavefunctions is that their coordinate and momentum space expressions are very similar. To see this, we Fourier-transform $\Phi^b_{nm}(x^\perp)$ to obtain the momentum space wavefunction $\tilde\Phi^b_{nm}(p^\perp)$,
\begin{align}
    \label{HO_wavefunction_fourier_transform}
    \nonumber \tilde\Phi^b_{nm}(p^\perp) &= \braket{p^\perp}{nm}=\int\! \ud^2 x^\perp\ e^{-i\vec{x}^\perp\cdot\vec{p}^\perp}\Phi_{nm}(x^\perp) \\
    &=(2\pi)\tilde f^b_{nm}(p)\tilde\chi_m(\phi)\; ,
\end{align}
in which
\begin{align}
    \tilde f^b_{nm}(p)=&\frac{\sqrt{2}}{b}\sqrt{\frac{n!}{(n+|m|)!}}e^{-p^2/(2b^2)} \left(\frac{p}{b}\right)^{|m|}L^{|m|}_n\left(\frac{p^2}{b^2}\right)\ ,
\end{align}
and
\begin{align}
    \tilde\chi_m(\phi)=\frac{1}{\sqrt{2\pi}}e^{im\phi}\ .
\end{align}
The coordinate and momentum space wavefunctions (\ref{transverse_HO_wavefunction}) and (\ref{HO_wavefunction_fourier_transform}) differ only in an overall coefficient and in that the 2D-HO parameter $b$ appears in numerators or denominators, respectively. Note in particular that the wavefunctions depend only on $b$, not on $M$ and $\Omega$ individually.

\subsection{Basis truncation: IR and UV cutoffs}\label{CUTOFF}
As discussed in the paper, the BLFQ basis must be truncated in order for numerical calculations to be feasible. One of the conditions used for obtaining a finite dimensional basis space is that the transverse degrees of freedom of the multi-particle states obeys
\be
	\sum_\text{particles} 2n_l+| m_l |+1 \leq N_\text{max} \;.
\ee
This restriction also imposes both IR and UV cutoffs into our theory, as can be seen from the behavior of the 2D-HO wavefunctions.

The {\it momentum} space HO wavefunction~(\ref{HO_wavefunction_fourier_transform}) exhibits a sharp fall-off at around $p_\perp\sim 2b \sqrt{2n+|m|}$. The maximal transverse momentum can be supported by the basis spaces truncated at $N_\text{max}$ is therefore around $p^\text{max}_\perp\propto b\sqrt{N_\text{max}}$, which is an {\it ultraviolet} cutoff in momentum space.

Since the coordinate space wavefunctions are so closely related to those in momentum space, we see immediately that the same basis states in coordinate space have support up to $x^\text{max}_\perp\propto \sqrt{N_\text{max}}/b$. This translates into an {\it infrared} cutoff in the momentum space as $p^\text{min}_\perp=1/x^\text{max}_\perp\propto b/\sqrt{N_\text{max}}$. The above UV and IR cutoffs are analogous to cutoffs of the 3D-HO that have recently been analyzed in low-energy nuclear physics applications \cite{Coon:2012ab,Furnstahl:2012qg}.

\section{Light-front QED in the BLFQ basis}
\label{OpBLFQbasis_BLFQ_basics}
%
\subsection{Mode expansion}
The mode expansion for field operators in the BLFQ basis is,
\begin{align}
    \label{field_op_momentum_basis_BLFQ}
    \Psi(x) &= \sum_{\bar\alpha}\frac{1}{\sqrt{2L}} \int\! \frac{\ud^2p^\perp}{(2\pi)^2} \big[b_{\bar\alpha}\tilde\Phi_{nm}(p^\perp)u(p,\lambda) e^{-i{\sf p\cdot x}} \nonumber \\
    &\qquad +d^\dagger_{\bar\alpha} \tilde\Phi^*_{nm}(p^\perp)v(p,\lambda) e^{i{\sf p\cdot x}} \big]  \;, \\
    A_\mu(x)&=\sum_{\bar\alpha}\frac{1}{\sqrt{2Lp^+}} \int\! \frac{\ud^2p^\perp}{(2\pi)^2} \big[ a_{\bar\alpha}\tilde\Phi_{nm}(p^\perp)\epsilon_\mu(p,\lambda)e^{-i{\sf p\cdot x}}  \nonumber \\
     &\qquad +a^\dagger_{\bar\alpha}\tilde\Phi^*_{nm}(p^\perp)\epsilon^*_\mu(p,\lambda)e^{i{\sf p\cdot x}}\big] \;,
\end{align}
where ${\sf p\cdot x}=\frac{1}{2}p^+x^--p^\perp\cdot x^\perp$ is the 3-product for the spatial components of $p^\mu$ and $x^\mu$, and see Eq.~(\ref{longitudinal_momentum}) for the values of $p^\LCp$, which depend on the (anti-)periodic boundary conditions for (fermions) gauge bosons. The creation operators $b^\dagger_{\bar\alpha}$, $d^\dagger_{\bar\alpha}$ and $a^\dagger_{\bar\alpha}$ create electrons, positrons and photons (respectively) with quantum numbers $\bar\alpha=\{k,n,m,\lambda\}$. They obey the (anti-)commutation relations
\be
	 [a_{\bar\alpha},a^\dagger_{{{\bar\alpha}'}}] = \{ b_{\bar\alpha},b^\dagger_{{\bar\alpha}'} \} = \{ d_{\bar\alpha},d^\dagger_{{\bar\alpha}'} \} = \delta_{\bar{\alpha}\bar{\alpha}'} \,.
\ee
With this, and using the explicit forms of the spinors and polarization vectors, given below, one can verify that the fields obey the standard equal-light-front-time commutation relations,
\begin{align}
    \label{communation_field_op}
    \nonumber &\big\{\Psi_+(x),\Psi^\dagger_+(y)\big\}_{x^+=y^+} = \Lambda^+\delta(x^--y^-)\delta^2(x^\perp-y^\perp)\ ,\\
    &\big[ A_i(x),A_j(y) \big]_{x^+=y^+} = \frac{-i}{4}\delta_{ij}\epsilon(x^--y^-)\delta^2(x^\perp-y^\perp)\;,
\end{align}
in which $\Lambda^\LCpm = \gamma^\LCmp \gamma^\LCpm / 4$ are the usual orthogonal, lightfront projectors and $\epsilon(x)$ is the sign function.

\subsection{Spin and polarization}

We use the following (chiral) spinor representation, with helicity $\lambda=\pm 1/2=\uparrow\downarrow$
\begin{align}
    &u(p,\uparrow)=\left(\begin{matrix}1\cr 0\cr \frac{im_e}{p^+}\cr  
\frac{(ip^1-p^2)}{p^+}\cr\end{matrix}\right), \quad
    u(p,\downarrow)=\left(\begin{matrix}0\cr 1\cr \frac{(-ip^1-p^2)}{p^+}\cr \frac{im_e}{p^+}\cr  
\end{matrix}\right), \quad\nonumber\\
    &v(p,\uparrow)=\left(\begin{matrix}0\cr 1\cr\frac{(-ip^1-p^2)}{p^+}\cr\frac{-im_e}{p^+}\cr\end{matrix}\right), \quad
    v(p,\downarrow)=\left(\begin{matrix}1\cr 0\cr  \frac{-im_e}{p^+}\cr  \frac{(ip^1-p^2)}{p^+}\cr
\end{matrix}\right).
\end{align}
We use a circularly polarized basis of polarization vectors for the photon, with $\lambda=\pm 1=\uparrow\downarrow$,
\begin{align}
\epsilon^\mu(k,\lambda)=\left(0,\epsilon^\perp(\lambda),
\frac{2\epsilon^\perp(\lambda)\cdot k^\perp}{k^+}\right),
\label{gluonpol}
\end{align}
in which the transversal polarization vectors are $\epsilon^{\perp}(+1)=\frac{1}{\sqrt{2}}(1,i)$ and  $\epsilon^{\perp}(-1)=\frac{1}{\sqrt{2}}(1,-i)$. The vectors are normalized according to
\begin{align}
\epsilon^\mu(k,\lambda)\epsilon^\ast_\mu(k,\lambda^\prime) =-\delta_{\lambda\lambda^\prime} \;. 
\end{align}

\subsection{The Hamiltonian}

We have now written our free field operators in terms of creation and annihilation operators. The next step is therefore to express the Hamiltonian in terms of the same basis. Hence, we take the operators (\ref{field_op_momentum_basis_BLFQ}) and insert them into (\ref{FULL}). The calculation is lengthy and unenlightening, so we will simply give some example terms.

First, the kinetic energy term for the fermions, which is the first term of (\ref{FULL}) : 
\begin{widetext}
\begin{align} 
    \label{QED_Hami_kinetic_BLFQ_Te}
  \nonumber  T_f=\int^L_{-L}\!\ud & x^-\!\! \int\! \ud^2 x^\perp \frac{1}{2}\bar{{\Psi}} \gamma^\LCp \frac{m^2_e+(i\partial^\LCperp)^2}{i\partial^\LCp}\Psi  =\sum_{\bar\alpha \bar\alpha'}\frac{1}{p^+} (b^\dagger_{\bar\alpha'}b_{\bar\alpha}+d^\dagger_{\bar\alpha'}d_{\bar\alpha}) \times \\
    &\times \bigg([m^2_e+(2n+|m|+1)b^2]\delta^{\bar\alpha'}_{\bar\alpha} -b^2[\sqrt{(n+1)(n+|m|+1)}\delta^{n'-1}_{n}+\sqrt{n(n+|m|)}\delta^{n'+1}_{n}]  \delta^{m'}_{m}\delta^{k'}_{k}\delta^{\lambda'}_{\lambda}\bigg) \;.
\end{align}
\end{widetext}
One can similarly obtain expressions for the interaction terms in (\ref{field_op_momentum_basis_BLFQ}). The vertex interaction $W_1$ becomes, for example
\begin{widetext}
\begin{align}
    \label{QED_Hami_vertex_BLFQ}
    W_1&=e\int^L_{-L}\!\ud x^-\!\! \int\!\ud^2x^\perp \bar\Psi\gamma^\mu\Psi A_\mu =\frac{e}{(2\pi)^4\sqrt{2L}}\sum_{\bar\alpha_1\bar\alpha_2\bar\alpha_3} \frac{1}{\sqrt{p^+_2}} \int\! \ud^2(p^\perp_{1},p^\perp_{2},p^\perp_{3}) \times \nonumber \\
    &\qquad \times \big[\tilde\Phi_{n_1 m_1}(p^\perp_1)\tilde\Phi^*_{n_2 m_2}(p^\perp_2)\tilde\Phi^*_{n_3 m_3}(p^\perp_3)\bar{u}(p_3,\lambda_3)\gamma^\mu\epsilon^*_\mu(p_2,\lambda_2)u(p_1,\lambda_1)\delta^{(3)}(p_1-p_2-p_3)b^\dagger_{\bar\alpha_3}a^\dagger_{\bar\alpha_2}b_{\bar\alpha_1} \nonumber \\ 
    &\qquad\quad +\tilde\Phi_{n_1m_1}(p^\perp_1)\tilde\Phi_{n_2m_2}(p^\perp_2)\tilde\Phi^*_{n_3m_3}(p^\perp_3)\bar{u}(p_3,\lambda_3)\gamma^\mu\epsilon_\mu(p_2,\lambda_2)u(p_1,\lambda_1)\delta^{(3)}(p_1+p_2-p_3)b^\dagger_{\bar\alpha_3}a_{\bar\alpha_2}b_{\bar\alpha_1}   \nonumber \\
    &\qquad\quad +\text{ six similar terms.} \big]
    \end{align}
\end{widetext}
Here $\bar\alpha_1,\bar\alpha_2,\bar\alpha_3$ are the quantum numbers associated with the field operators  $\Psi$, $A_\mu$ and $\bar\Psi$ respectively. The 3D-$\delta$ functions should be understood as the Dirac delta function for the transverse momentum ($p^\perp$) and the Kronecker delta for the discretized longitudinal momentum ($p^+$). The two terms given above cause transitions between the $\ket e$ and $\ket{e\gamma}$ Fock sectors. The six terms which we have not written explicitly do not contribute to the calculations in this paper, as they describe transitions between Fock sectors which are not present in our truncated basis space. The spinor-polarization vector contraction part  $\bar{u}(p_3,\lambda_3)\gamma^\mu\epsilon^*_\mu(p_2,\lambda_2)u(p_1,\lambda_1)$ in the first term is summarized in Table~\ref{tab:spinor-polarization_contraction_list} for different helicity configurations. Taking complex conjugates and changing labels gives the results with $\epsilon_\mu$ instead for $\epsilon_\mu^*$. Integration over the product of three, highly oscillatory, 2D-HO wavefunctions, as in (\ref{QED_Hami_vertex_BLFQ}), would pose a challenge for numerical calculations. Fortunately, this type of integral can be performed analytically by applying the Talmi-Moshinsky transformation to the 2D-HO wavefunctions, see~\cite{Davies}.
\begin{table}[ht!]
    \renewcommand{\arraystretch}{2.5}
    \centering
\begin{tabular}{|c|c|}
\hline
helicity config. ($\lambda_3$,$\lambda_2$,$\lambda_1$) & $\bar{u}(p_3,\lambda_3)\gamma^\mu\epsilon^*_\mu(p_2,\lambda_2)u(p_1,\lambda_1)$ \\ \hline
$\uparrow\uparrow\uparrow$ & $-\sqrt{2}\frac{p^1_3-ip^2_3}{p^+_3}+\sqrt{2}\frac{p^1_2-ip^2_2}{p^+_2}$\\ \hline
$\uparrow\uparrow\downarrow$ & 0\\ \hline
$\uparrow\downarrow\uparrow$ & $\sqrt{2}\frac{p^1_2+ip^2_2}{p^+_2}-\sqrt{2}\frac{p^1_1+ip^2_1}{p^+_1}$\\ \hline
$\uparrow\downarrow\downarrow$ & $\sqrt{2}\frac{m_e}{p^+_3}-\sqrt{2}\frac{m_e}{p^+_1}$\\ \hline
$\downarrow\uparrow\uparrow$ & $-\sqrt{2}\frac{m_e}{p^+_3}+\sqrt{2}\frac{m_e}{p^+_1}$\\ \hline
$\downarrow\uparrow\downarrow$ & $\sqrt{2}\frac{p^1_2-ip^2_2}{p^+_2}-\sqrt{2}\frac{p^1_1-ip^2_1}{p^+_1}$\\ \hline
$\downarrow\downarrow\uparrow$ & 0\\ \hline
$\downarrow\downarrow\downarrow$ & $-\sqrt{2}\frac{p^1_3+ip^2_3}{p^+_3}+\sqrt{2}\frac{p^1_2+ip^2_2}{p^+_2}$\\ \hline
\end{tabular}
\caption{Spinor-polarization vector contraction for different helicity configurations of the incoming electron (``1''), outgoing photon (``2'') and the outgoing electron (``3'').}
\label{tab:spinor-polarization_contraction_list}
\end{table}
\section{Constructing tBLFQ basis: An example}
\label{sample_BLFQ}
%
In this Appendix we illustrate the construction of the extended BLFQ basis, the diagonalization of the Hamiltonian and the construction of the tBLFQ basis. For the sake of clarity we work in a highly truncated basis space for which numerical results are subject to large truncation error, but this section is for illustration only.

\subsection{BLFQ state enumeration}
We take $b{=}m_e$ and $L{=}2\pi$\,MeV$^{-1}$. In this example, our extended BLFQ basis consists of two segments, labeled by $\{K{=}3/2, M_j{=}1/2, N_f{=}1\}$ and $\{K{=}5/2, M_j{=}1/2, N_f{=}1\}$. In each segment we truncate the transverse degrees of freedom at $N_\text{max}{=}2$.

Consider now which Fock states are present in our basis. In each segment, the basis states have to meet the symmetry constraints (\ref{basis_symconstrain_k}) to~(\ref{basis_symconstrain_nf}), and the truncation constraint~(\ref{Nmax_truncation}). Consequently, there are only two states in the $K{=}3/2$ segment; their quantum numbers are given in Table~\ref{tab:BLFQ_state_list_K1_segment}. In the $K{=}5/2$ segment we have 3 basis states, see Table~\ref{tab:BLFQ_state_list_K2_segment}. The total dimensionality of this extended BLFQ space, which is simply the sum of the two segments, is 2+3=5.
\begin{table}[ht!]
    \renewcommand{\arraystretch}{1.5}
    \centering
\begin{tabular}{|c|c|c|c|c|c|c|c|c|c|}
\hline
basis state no.  & Fock-sector & $k^e$ & $n^e$ & $m^e$ & $\lambda^e$ & $k^\gamma$ & $n^\gamma$ & $m^\gamma$ & $\lambda^\gamma$\\ \hline
1 & $\ket{e}$ & 3/2 & 0 & 0 & 1/2 & - & - & - & - \\ \hline
2 & $\ket{e\gamma}$ & 1/2 & 0 & 0 & -1/2 & 1 & 0 & 0 & 1 \\ \hline
\end{tabular}
\caption{\label{tab:BLFQ_state_list_K1_segment} BLFQ basis states in the segment \{$K{=}3/2$, $M_j{=}1/2$, $N_f{=}1$\}.}
\end{table}
\begin{table}[ht!]
    \renewcommand{\arraystretch}{1.5}
    \centering
\begin{tabular}{|c|c|c|c|c|c|c|c|c|c|}
\hline
basis state no.  & Fock-sector & $k^e$ & $n^e$ & $m^e$ & $\lambda^e$ & $k^\gamma$ & $n^\gamma$ & $m^\gamma$ & $\lambda^\gamma$\\ \hline
3 & $\ket{e}$ & 5/2 & 0 & 0 & 1/2 & - & - & - & - \\ \hline
4 & $\ket{e\gamma}$ & 3/2 & 0 & 0 & -1/2 & 1 & 0 & 0 & 1 \\ \hline
5 & $\ket{e\gamma}$ & 1/2 & 0 & 0 & -1/2 & 2 & 0 & 0 & 1 \\ \hline
\end{tabular}
\caption{\label{tab:BLFQ_state_list_K2_segment} BLFQ basis states in the segment of \{$K{=}5/2$, $M_j{=}1/2$, $N_f{=}1$\}.}
\end{table}
Note that we have assigned a number to each of the basis states; this ordering is a matter of choice, and we include it so that the structures of the exact QED eigenstates which we will construct below, can be more easily related to their Fock components.

Now that we have the basis states $\ket{\alpha}$, we calculate their matrix elements with the Hamiltonian, $\bra{\alpha'} P^-_\text{QED}\ket{\alpha}$, using for example the expressions (\ref{QED_Hami_kinetic_BLFQ_Te}) and (\ref{QED_Hami_vertex_BLFQ}). The resulting QED Hamiltonian in our 5-dimensional extended BLFQ basis is shown in Table~\ref{tab:QED_Hamiltonian}.
\begin{table}[tbh!]
    \renewcommand{\arraystretch}{1.5}
    \centering\begin{tabular}{|ccc|c|c|c|c|c|c|c|c|c|c|}
\hline
\multicolumn{3}{|c|}{$\bra{\alpha'}P^\LCm_\text{QED}\ket{\alpha}$} & \multicolumn{10}{c|}{BLFQ basis state $\ket{\alpha}$}\\ \cline{4-13}
  (MeV) & & & 1 & $\ket{e}$ & 2 & $\ket{e\gamma}$ & 3 & $\ket{e}$ & 4 & $\ket{e\gamma}$ & 5 & $\ket{e\gamma}$ \\ \hline
\multirow{5}{1cm}{BLFQ basis state $\bra{\alpha'}$} & \multicolumn{1}{|c}{1} & \multicolumn{1}{|l}{$\ket{e}$} & \multicolumn{2}{|c}{0.3482} & \multicolumn{2}{|c}{-0.0119} & \multicolumn{2}{|c}{0} & \multicolumn{2}{|c}{0} & \multicolumn{2}{|c|}{0} \\ \cline{2-13}
& \multicolumn{1}{|c}{2} & \multicolumn{1}{|l}{$\ket{e\gamma}$} & \multicolumn{2}{|c}{-0.0119} & \multicolumn{2}{|c}{0.9139} & \multicolumn{2}{|c}{0} & \multicolumn{2}{|c}{0} & \multicolumn{2}{|c|}{0} \\ \cline{2-13}
& \multicolumn{1}{|c}{3} & \multicolumn{1}{|l}{$\ket{e}$} & \multicolumn{2}{|c}{0} & \multicolumn{2}{|c}{0} & \multicolumn{2}{|c}{0.2089} & \multicolumn{2}{|c}{-0.0024} & \multicolumn{2}{|c|}{-0.0101} \\ \cline{2-13}
& \multicolumn{1}{|c}{4} & \multicolumn{1}{|l}{$\ket{e\gamma}$}& \multicolumn{2}{|c}{0} & \multicolumn{2}{|c}{0} & \multicolumn{2}{|c}{-0.0024} & \multicolumn{2}{|c}{0.3917} & \multicolumn{2}{|c|}{0} \\ \cline{2-13}
& \multicolumn{1}{|c}{5} & \multicolumn{1}{|l}{$\ket{e\gamma}$} & \multicolumn{2}{|c}{0} & \multicolumn{2}{|c}{0} & \multicolumn{2}{|c}{-0.0101} & \multicolumn{2}{|c}{0} & \multicolumn{2}{|c|}{0.8486} \\ \hline
\end{tabular}
\caption{\label{tab:QED_Hamiltonian} The Hamiltonian matrix $P^-_\text{QED}$ in the extended BLFQ basis consisting of two segments: $\{K{=}3/2,M_j{=}1/2,N_f{=}1\}$ and $\{K{=}5/2,M_j{=}1/2,N_f{=}1\}$. See text for truncation parameters.}
\end{table}
The diagonal entries come from the free kinetic terms in the Hamiltonian. (In general, the kinetic terms give off-diagonal matrix elements, as can be seen from (\ref{QED_Hami_kinetic_BLFQ_Te}); such terms do not appear here only because of the small basis space.) The off-diagonal matrix elements in Table~\ref{tab:QED_Hamiltonian} come from the vertex interaction $W_1$. As expected, the Hamiltonian matrix exhibits a block-diagonal structure; no coupling exists between the $K{=}3/2$ segment (states 1 and 2) and the $K{=}5/2$ segment (states 3, 4 and 5), due to the symmetries of QED. Normally, we would also include the instantaneous $W_3$ terms. However, as noted above, we are only solving for QED mass eigenstates with the Hamiltonian accurate to order $e$ in the present work.  
\begin{widetext}

\subsection{Diagonalization of $P^-_\text{QED}$}
\begin{table*}[h!]
    \renewcommand{\arraystretch}{1.5}
    \centering
\begin{tabular}{|c|c|c|c||c|c|c|c|c|}
\hline
$P^-_\text{QED}$ eigenstate & \multirow{2}{*}{$P^-_\beta$(MeV)} & \multirow{2}{*}{$M_\beta$(MeV)} & & \multicolumn{5}{c|}{BLFQ amplitudes $\braket{\beta}{\alpha}$} \\ \cline{4-9}
$\bra{\beta}$ & & & $K$ & 1\,$\ket{e}$ & 2\,$\ket{e\gamma}$ & 3\,$\ket{e}$ & 4\,$\ket{e\gamma}$ & 5\,$\ket{e\gamma}$ \\ \hline
1 & 0.3479 & 0.5106 & 3/2  & -0.9998 & -0.0210 & 0 & 0 & 0 \\ \hline
2 & 0.9142 & 1.0540 & 3/2 & -0.0210 & 0.9998 & 0 & 0 & 0 \\ \hline
3 & 0.2087 & 0.5105 & 5/2 & 0 & 0 & 0.9998 & 0.0130 & 0.0157 \\ \hline
4 & 0.3917 & 0.8474 & 5/2 & 0 & 0 & -0.0130 & 0.9999 & -0.0003 \\  \hline
5 & 0.8488 & 1.3640 & 5/2 & 0 & 0 & -0.0157 & 0.0001 & 0.9999 \\ \hline
\end{tabular}
\caption{\label{tab:eigenspectrum_QED_Hamiltonian} Eigenstates and eigenvalues of the QED Hamiltonian in the extended BLFQ basis comprising the two segments $\{K{=}3/2,M_j{=}1/2,N_f{=}1\}$ and $\{K{=}5/2,M_j{=}1/2,N_f{=}1\}$.}
\end{table*}

With the QED Hamiltonian matrix in the BLFQ basis prepared, we are ready to diagonalize it. This can be done segment by segment because of the block-diagonal structure of the Hamiltonian. Doing so, we obtain eigenstates $\ket{\beta}$ and eigenvalues $P^-_\beta$. These are listed in Table.~\ref{tab:eigenspectrum_QED_Hamiltonian}. The first column enumerates the eigenstates. The second and third columns contain $P^-_\beta$ and the invariant mass $M_\beta$ for each eigenstate, which will be discussed shortly. The fourth column contains the segment specifier; it is enough to give just $K$ in this case.   In the 5th to 9th columns we list the overlaps of the eigenstates $\ket{\beta}$ with the BLFQ basis states $\ket{\alpha}$ in Tables~\ref{tab:BLFQ_state_list_K1_segment} and~\ref{tab:BLFQ_state_list_K2_segment}, \ie\ this part of the table contains the coefficients in the expansion
\be\label{alpha-beta}
	\ket\beta = \sum_\alpha \ket{\alpha}\braket{\alpha}{\beta} \;.
\ee
Let us comment briefly on the physical interpretations of these QED eigenstates. In this example, the five QED eigenstates lie in two segments. The eigenstates numbered 1 and 2 are in the (total longitudinal momentum) $K=3/2$ segment, while eigenstates 3, 4 and 5 are in the $K=5/2$ segment.  In order to interpret these states it is useful to introduce the invariant mass,
\be\label{inv_mass}
	M^2 := P^\LCp P^\LCm -P^\LCperp P^\LCperp \;.
\ee
We see  that the eigenstates 1 and 3 have invariant masses close to the physical electron mass $m_e$. Reading off the coefficients (\ref{alpha-beta}), we see that these states are dominated by contributions from the single electron basis states. Thus we interpret them as the physical single electron states $\ket{e_\text{phys}}$ with different longitudinal momenta. (The slight deviation in the invariant mass from $m_e$ is due to our omission of mass corrections from counter-terms.) The small $\ket{e\gamma}$ components in their wavefunctions are generated by the QED vertex interaction ($W_1$), and describe the dressing of the bare fields by the photon cloud which, together, make up the physical electron \cite{Dirac:1955uv,Lavelle:1995ty,Bagan:1999jf}. We will see that these $\ket{e\gamma}$ components play an important role in photon-radiation processes. They are also responsible for the electron's anomalous magnetic moment, see Ref.~\cite{Zhao:2013xx} for more details. 

The eigenstates 2, 4 and 5 are excited states in their respective segments, with invariant masses considerably above the physical electron mass. Since they are dominated by the basis states in the $\ket{e\gamma}$ sector, it is natural to interpret them as the electron-photon scattering states $\ket{e\gamma_\text{scat}}$. Their invariant masses $M_\beta=\sqrt{(P_e+P_\gamma)^2}$ are experimentally accessible through simultaneous measurements of the electron and photon four momenta. The $\ket{e\gamma_\text{scat}}$ states receive small contributions from the single Fock electron sector ($\ket{e}$) due to the QED vertex interaction $W_1$. As we will see below, it is through such ``minor'' components that external fields are able to couple physical electron states to electron-photon scattering states or, in other words, cause photon emission.

\subsection{tBLFQ: time evolution}
\label{sample_tBLFQ}
The eigenstates we have constructed comprise the tBLFQ basis. We continue our example by calculating transitions between these eigenstates in nCs.  The laser profile used in this example is
\begin{align}
    \label{laser-profile_one_exponential_example}
    e\mathcal{A}^-(x^-) &= 2a_0m_e\cos{(l_\LCm x^-)} = a_0 m_e \left[\exp{(il_\LCm x^-)}+\exp{(-il_\LCm x^-)}\right]\;,
\end{align}
and we take $a_0{=}1$, which is at the edge of the nonperturbative intensity regime. Recalling that the frequency $l_\LCm$ can be written in terms of the wave number $k_\text{las}$ as $l_\LCm{=}\frac{\pi}{L}k_\text{las}$, we take $k_\text{las}{=}1$. The laser can therefore cause transitions between just the two segments of the tBLFQ basis prepared above.

We need the matrix elements of $V$ in the tBLFQ basis. One can first write down the matrix elements in the extended BLFQ basis, \ie\ the set $\bra{\alpha'}V\ket{\alpha}$, and then transform to the tBLFQ basis using
\begin{align}
    \bra{\beta'}V\ket{\beta}=\sum_{\alpha\alpha'}\braket{\beta'}{\alpha'}\bra{\alpha'}V\ket{\alpha}\braket{\alpha}{\beta}\ .
\end{align}
\begin{table*}[t!]
    \renewcommand{\arraystretch}{1.5}
    \centering
\begin{tabular}{|c|c|c|c|c|c|c|}
\hline
\multirow{2}{*}{$\bra{\beta'}V\ket{\beta}$} & \multicolumn{6}{c|}{Basis element $\ket{\beta}$} \\ \cline{2-7}
(MeV) & & 1 & 2 & 3 & 4 & 5 \\ \hline
\multirow{5}{1.2cm}{Basis element $\bra{\beta'}$} &1 ($K$=3/2) & 0 & 0 & -0.5109 & -0.0041 & 0.0080 \\ \cline{2-7}
& 2 ($K$=3/2) & 0 & 0 & -0.0041 & 0.5110 & 0.0002 \\ \cline{2-7}
& 3 ($K$=5/2) & -0.5109 & -0.0041 & 0 & 0 & 0 \\ \cline{2-7}
& 4 ($K$=5/2) & -0.0041 & 0.5110 & 0 & 0 &0 \\ \cline{2-7}
& 5 ($K$=5/2) & 0.0080 & 0.0002 & 0 & 0 & 0 \\ \hline
\end{tabular}
\caption{\label{tab:laser_matrix_tBLFQ} The matrix elements of the interaction term $V$ in the tBLFQ basis. }
\end{table*}
The resulting matrix of $V$ in the tBLFQ basis, $\bra{\beta'}V\ket{\beta}$, is shown in Table.~\ref{tab:laser_matrix_tBLFQ}. The only allowed transitions are now {\it between} the two segments, because of the longitudinal momentum being added to the system by the background.  The most probable transitions are those between the physical electron states $\ket{e_\text{phys}}$ in the two segments, and between the electron-photon scattering states $\ket{e\gamma_\text{scat}}$ in the two segments. These types of transitions describe {\it acceleration}, as the particle number is conserved, but the longitudinal momentum is changed by one unit.

There are also transitions between the physical electron states and the electron-photon scattering states ($\ket{e_\text{phys}}\leftrightarrow\ket{e\gamma_\text{scat}}$), which describe the {\it radiation} process. We see that these transitions have much smaller amplitudes, since they link the ``minor'' Fock components in tBLFQ basis states $\ket{\beta}$ (\eg, the $\ket{e\gamma}$ components in $\ket{e_\text{phys}}$, and the $\ket{e}$ components in $\ket{e\gamma_\text{scat}}$), which are suppressed by one factor of the electron charge $e$. Next we multiply by the required phase factor $e^{i\omega_{\beta'\beta}x^+/2}$, which transforms the matrix elements into those in the interaction picture,
\begin{align}
    \label{V^I_L_interact}
  \bra{\beta'}V_I(x^+)\ket{\beta}= \bra{\beta'}V\ket{\beta} \times e^{i\omega_{\beta'\beta}x^+/2}\ ,
\end{align}
where $\omega_{\beta'\beta}=P^-_{\beta'}-P^-_\beta$ and (see Table.~\ref{tab:eigenspectrum_QED_Hamiltonian} for the values of these energy eigenvalues). Due to this phase factor, the transition amplitudes oscillate in time with the period, ${\sim}1/\omega_{\beta'\beta}$, inversely proportional to the light-front energy difference between $\ket{\beta'}$ and $\ket{\beta}$. Thus in $x^+{\to}\infty$ limit only the transitions which conserve light-front energy can accumulate. The interaction picture matrix elements of $V_I$ are given in Table.~\ref{tab:laser_matrix_tBLFQ_interact}.
\begin{table}[h!]
    \renewcommand{\arraystretch}{1.5}
    \centering
\begin{tabular}{|c|cc|c|c|c|c|c|}
\hline
\multicolumn{3}{|c|}{$\bra{\beta'}V_I(x^+)\ket{\beta}$ (MeV)} & \multicolumn{5}{|c|}{tBLFQ basis state $\ket{\beta}$} \\ \cline{4-8}
  \multicolumn{3}{|c|}{} & 1 & 2 & 3 & 4 & 5 \\ \hline
  \multirow{5}{1.2cm}{tBLFQ basis state $\ket{\beta'}$} & &1 ($K$=3/2)& 0 & 0 & $-0.5109e^{0.070ix^+}$ & $-0.0040e^{-0.022ix^+}$ & $0.0080e^{-0.250ix^+}$ \\ \cline{2-8}
& & 2 ($K$=3/2)& 0 & 0 & $-0.0040e^{0.353ix^+}$ & $0.5110e^{0.261ix^+}$ & $0.0002e^{0.033ix^+}$ \\ \cline{2-8}
& & 3 ($K$=5/2)& $-0.5109e^{-0.070ix^+}$ & $-0.0040e^{-0.353ix^+}$ &0 & 0 & 0 \\ \cline{2-8}
& & 4 ($K$=5/2)& $-0.0040e^{0.022ix^+}$ & $0.5110e^{-0.261ix^+}$  & 0 & 0 & 0\\ \cline{2-8}
& & 5 ($K$=5/2)& $0.0080e^{0.250ix^+}$ & $0.0002e^{-0.033ix^+}$ & 0 & 0 & 0\\ \hline
\end{tabular}
\caption{The matrix elements of the interaction term $V_I$ in the interaction picture. $x^\LCp$ is in units of MeV$^{-1}$. }
\label{tab:laser_matrix_tBLFQ_interact}
\end{table}
We also need the initial state of the system, $\ket{\psi;x^+ =0}_I$, cf. Eq.~(\ref{initial_c}). In the nCs process the initial state is a physical electron. In our current tBLFQ basis there are two states corresponding to physical electrons, the two $\ket{e_\text{phys}}$ states, see Tables~\ref{tab:BLFQ_state_list_K1_segment} and \ref{tab:BLFQ_state_list_K2_segment}. We choose our initial state to be that in the $K=3/2$ segment. Now we are in a position to evolve the initial state forward in $x^+$. To do so, we must identify the largest (by magnitude) eigenvalue of $V_I$ in order to determine our step size (see the discussion in Sect.~\ref{NumericalScheme}). According to  Table.~\ref{tab:laser_matrix_tBLFQ_interact} the smallest value for $1/\omega_{\beta'\beta}$ is about $\frac{1}{0.35}\sim$3MeV$^{-1}$, and according to Table.~\ref{tab:laser_matrix_tBLFQ} the largest eigenvalue (by magnitude) of $V_I$ is about 0.511MeV which translates to $1/|V_{I;\text{max}}|\sim$2MeV$^{-1}$. For this problem, we can safely choose the step size of $\delta x^+{=}0.1$MeV$^{-1}$.

\begin{table}[hbp]
    \renewcommand{\arraystretch}{1.5}
    \centering
\begin{tabular}{|cc|c|c|c|c|c|c|c|c|c|}
\hline
\multicolumn{2}{|c|}{\multirow{2}{*}{$|c_\beta|^2$}} & \multicolumn{9}{|c|}{$x^+$ (MeV$^{-1}$) }\\ \cline{3-11}
 & & 0 & 0.05 & 0.1 & 0.2 & 1.0 & 5.0 & 10.0 & 15.0 & 20.0\\ \hline
 \multicolumn{1}{|c}{1} & \multicolumn{1}{c|}{($K$=3/2)} & 1.000000 & 1.000000 & 0.999347 & 0.997391 & 0.936139 & 0.093892 & 0.720013 & 0.566033 & 0.199441\\ \cline{1-11}
 \multicolumn{1}{|c}{2} & \multicolumn{1}{c|}{($K$=3/2)} & 0.000000 & 0.000000 & 0.000000 & 0.000000 & 0.000000 & 0.000001 & 0.000010 & 0.000005 & 0.000005 \\ \cline{1-11}
 \multicolumn{1}{|c}{3} & \multicolumn{1}{c|}{($K$=5/2)} & 0.000000 & 0.000163 & 0.000653 & 0.002609 & 0.063841 & 0.905839 & 0.279520 & 0.432517 & 0.798676\\ \cline{1-11}
 \multicolumn{1}{|c}{4} & \multicolumn{1}{c|}{($K$=5/2)} & 0.000000 & 0.000000 & 0.000000 & 0.000000 & 0.000004 & 0.000060 & 0.000035 & 0.000005 & 0.000015\\ \cline{1-11}
 \multicolumn{1}{|c}{5} & \multicolumn{1}{c|}{($K$=5/2)} & 0.000000 & 0.000000 & 0.000000 & 0.000001 & 0.000016 & 0.000209 & 0.000422 & 0.001440 & 0.001862\\ \hline
\multicolumn{2}{|c|}{$|\braket{\psi}{\psi}|^2$} & 1.000000 & 1.000000 & 1.000000 & 1.000000 & 1.000000 & 1.000000 & 1.000000 & 1.000000 & 0.999999\\ \hline
\multicolumn{2}{|c|}{$\langle M_\psi\rangle$ (MeV)} & 0.510635 & 0.510635 & 0.510635 & 0.510635 & 0.510643 & 0.510745 & 0.510985 & 0.511826 & 0.512153 \\
\hline
\end{tabular}
\caption{The evolution of $\ket{e_\text{phys}}$ with $K{=}3/2$ in the laser field by Eq.~(\ref{laser-profile_one_exponential_example}). The norm and invariant mass are listed in the 6th and 7th row, respectively.}
\label{tab:evolved_state_interact}
\end{table}

Starting from the initial state $\ket{\psi;0}_I$, \ie\ a physical electron with $K=3/2$, the evolved state $\ket{\psi;x^+}_I$ is represented in Table.~\ref{tab:evolved_state_interact}, at various times by the probabilities for being in various QED eigenstates $\ket{\beta}$. As time begins to evolve, the state first acquires an overlap with the $\ket{e_\text{phys}}$ state with $K = 5/2$ (the state with $\{\beta,K\}=\{3,5/2\}$).  Mathematically, this state is populated first due to the large matrix element in $V_I$ coupling it to the initial state. Physically, this is acceleration; the electron is accelerated by the laser but does not, yet, have a significant probability for photon emission.

At later times, the $\ket{e\gamma_\text{scat}}$ states ($\{\beta,K\} {=} \{4,5/2\}$ and $\{5,5/2\}$) become populated. The overlap between these states and the initial state is smaller than between the physical electron state with the initial state. Hence, we observe a ``domination'' of acceleration over radiation, as the latter is suppressed by a factor of the coupling. Note that the $\ket{e\gamma_\text{scat}}$ scattering state in the $K=3/2$ segment,  ($\{\beta, K\} {=}\{2,3/2\}$) eventually becomes populated, even though it is not directly coupled to the initial state. This population arises through the basis states in the $K {=} 5/2$ segment, which are ``decelerated" to the $\{\beta,K\}{=}\{2,3/2\}$ state by the laser field. (This is due to the presence of the ``negative'' exponential in the chosen laser profile (\ref{laser-profile_one_exponential_example}), which subtracts rather than adds longitudinal momentum.)

The norm of the state $\ket{\psi;x^+}$ is conserved over the entire evolution due to the stability of the MSD2 scheme. It is an advantage of the tBLFQ approach that the wavefunction of the system is accessible at each time step, which allows one to monitor the real-time evolution of any observable $O(x^+)$, by evaluating $O(x^+){=}{}_I\bra{\psi;x^+}\hat O_I\ket{\psi;x^+}_I$. For example, taking $\hat O_I$ to be the invariant mass operator, we obtain the evolution of the average invariant mass of the system, see the 7th row of Table.~\ref{tab:evolved_state_interact}. The increase in the invariant mass with time reflects the fact that the laser field pumps energy into the system and hence that photons are being created.

This completes our example. Performing calculations in larger basis spaces follows the same procedure. In Sec.~\ref{ssec:nonpert} we present the numerical results for the nCs process in larger basis spaces, which allows for a more accurate description of the evolution of the system.

\section{Comparison with perturbation theory}
\label{laserme_comp}
In order to check the BLFQ calculation, we calculate the matrix element $\bra{e\gamma_\text{scat}} V \ket{e_\text{phys}}$ in perturbation theory. As discussed in Sect.~\ref{ssec:cali}, the perturbative approximation to this matrix element follows from standard time-independent perturbation theory, and is equal to 
\be\begin{split}\label{near}
	\bra{e\gamma_\text{scat}} V \ket{e_\text{phys}} &= \bra{e\gamma} V_Q \frac{1}{P^\LCm_\text{free}(e\gamma)-\hat{P}^\LCm_\text{free}}V\ket{e} +\bra{e\gamma} V \frac{1}{P^\LCm_\text{free}(e)-\hat{P}^\LCm_\text{free}}V_Q\ket{e}\;.
\end{split}
\ee
The Feynman diagram representation of the perturbative matrix elements on the right hand side of (\ref{near}) is shown in Fig.~\ref{FF}; these are (respectively) $s$-channel and $t$-channel Compton scattering diagrams, where the laser field takes the place of the incoming photon. Using the machinery in Appendix~\ref{Transverse} to write these perturbative overlaps in terms of the BLFQ basis, one eventually finds (with $l^\LCp {=} 2 l_\LCm$ from (\ref{laser-profile}) and $\delta^*_*$ the Kronecker delta),
\begin{figure}[t!]
\includegraphics[width=0.3\columnwidth]{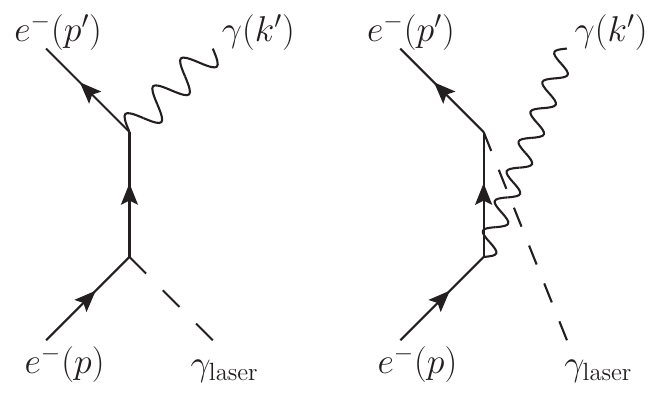}
\caption{\label{FF} Feynman diagram representation of the perturbative processes contributing to the matrix element of the background field vertex between exact QED eigenstates.}
\end{figure}

\be\begin{split}
	\bra{e\gamma} V_Q \frac{1}{P^\LCm_\text{free}(e\gamma)-\hat{P}^\LCm_\text{free}}V\ket{e} = \frac{e a_0 m_e }{4\sqrt{2 L}}\int\!\frac{\ud^2(p'^\LCperp,k'^\LCperp,p^\LCperp)}{(2\pi)^4{\sqrt{{k'}^\LCp}}}\tilde{\Phi}^*_{n''m''}(k'^\LCperp)\tilde{\Phi}^*_{n'm'}(p'^\LCperp)\tilde{\Phi}_{nm}(p^\LCperp) \\
	\sum\limits_{s=\pm 1} \delta^\LCperp(p'+k'-p)\delta^{p'^\LCp+k'^\LCp}_{p^\LCp+sl^\LCp} \frac{{p'}^\LCp + {k'}^\LCp}{k'.p'}	 \bar{u}_{\sf p'}^{\lambda'} \slashed{\epsilon}^*(k') u^\lambda_{{\sf p}+s{\sf l}} \;,
\end{split}
\ee
and
\be\begin{split}
	\bra{e\gamma} V \frac{1}{P^\LCm_\text{free}(e)-\hat{P}^\LCm_\text{free}}V_Q\ket{e} = -\frac{e a_0 m_e }{4\sqrt{2 L}}\int\!\frac{\ud^2(p'^\LCperp,k'^\LCperp,p^\LCperp)}{(2\pi)^4{\sqrt{{k'}^\LCp}}}\tilde{\Phi}^*_{n''m''}(k'^\LCperp)\tilde{\Phi}^*_{n'm'}(p'^\LCperp)\tilde{\Phi}_{nm}(p^\LCperp) \\
	\sum\limits_{s=\pm 1} \delta^\LCperp(p'+k'-p)\delta^{p'^\LCp+k'^\LCp}_{p^\LCp+sl^\LCp} \frac{p^\LCp - {k'}^\LCp}{k'.p}	 \bar{u}_{{\sf p'}-s {\sf l}^\LCp}^{\lambda'} \slashed{\epsilon}^*(k') u^\lambda_{{\sf p}} \;,
\end{split}
\ee
The final factors in each of the above, describing spin and polarization contributions, can be read off from Table~\ref{tab:spinor-polarization_contraction_list}.

\end{widetext}

\end{document}